\documentclass[preprint]{aastex}
\usepackage{graphicx}
\usepackage{natbib}

\shorttitle{Dust evolution in disks of Herbig AeBe stars}
\shortauthors{{Juh\'asz} et al.}

\begin{document}

\newcommand{\bface}{\bf }


\title{Dust evolution in protoplanetary disks around Herbig Ae/Be stars - The Spitzer view}


\author{A. Juh\'asz\altaffilmark{1}, J. Bouwman\altaffilmark{1}, Th. Henning\altaffilmark{1}, B. Acke\altaffilmark{2},
M.E. van den Ancker\altaffilmark{3}, G. Meeus\altaffilmark{4}, C. Dominik\altaffilmark{5}, M. Min\altaffilmark{4}, 
A.G.G.M. Tielens\altaffilmark{6,7}, L.B.F.M. Waters\altaffilmark{2,4}}

\affil{Max-Planck-Institut f\"ur Astronomie, K\"onigstuhl 17,  69117, Heidelberg, D69117 Germany}
\affil{Instituut voor Sterrenkunde, Katholieke Universiteit Leuven, Celestijnenlaan 200D, 3001 Leuven, Belgium}
\affil{European Southern Observatory, Karl Schwarzschild Strasse 2, 85748 Garching bei M\"unchen, Germany}
\affil{Astrophysical Institute Potsdam, An der Sternwarte 16, 14482 Potsdam, Germany}
\affil{Astronomical Institute, University of Amsterdam, Kruislaan 403, 1098 AJ Amsterdam, The Netherlands}
\affil{Kapteyn Astronomical Institute, University of Groningen, P.O. Box 800, 9700 AV, Groningen, Netherlands}
\affil{ NASA Ames Research Center, MS 245-3, Moffett Field, CA 94035, USA}



\begin{abstract}
In this paper we present mid-infrared spectra of a comprehensive set of Herbig Ae/Be stars observed with the Spitzer Space Telescope. 
The signal-to-noise ratio of these spectra is very high, ranging between about a hundred and several hundreds. During the analysis of 
these data we tested the validity of standard protoplanetary dust models and studied grain growth and crystal formation. 
On the basis of the analyzed spectra, the major constituents of protoplanetary dust around Herbig Ae/Be stars are
amorphous silicates with olivine and pyroxene stoichiometry, crystalline forsterite and enstatite and silica. 
No other solid state features, indicating other abundant dust species, are present in the Spitzer spectra.
Deviations of the synthetic spectra from the observations are most likely related to grain shape effects and uncertainties
in the iron content of the dust grains. 

Our analysis revealed that larger grains are more abundant in the disk atmosphere of flatter disks than in that of flared disks, indicating
that grain growth and sedimentation decrease the disk flaring. We did not find, however, correlations between the value of crystallinity
and any of the investigated system parameters. Our analysis shows that enstatite is more concentrated toward the warm inner disk
than forsterite, in contrast to predictions of equilibrium condensation models. 
None of the three crystal formation mechanisms proposed so far can alone explain all our findings. It is very likely
that all three play at least some role in the formation of crystalline silicates. 
\end{abstract}





\section{Introduction}

The class of Herbig Ae/Be (hereafter HAeBe) stars was established by \citet{ref:herbig1960} as
stars which are surrounded by nebulosities and the optical spectra of which show emission lines. 
Further investigations revealed that these sources are young stars (1--10\,Myr) 
with masses between 2 and 10\,M$_\odot$ in the later stages of their pre-main sequence evolution. 
Observations of these sources at infrared wavelengths revealed that excess emission above 
the stellar photosphere is another characteristic of HAeBe stars  (for a review, see \citet{ref:waters_waelkens1998}). 
The infrared excess emission arises from a protoplanetary disk \citep{ref:waters_waelkens1998} and 
in many cases from an envelope as well  \citep{ref:leinert2001}. 
HAeBe stars are, therefore, frequently regarded as the higher mass counterparts of the low-mass 
T Tauri stars. Planet formation theories suggest that this evolutionary stage (1--10\,Myr), 
is exactly where the formation of planetary embryos is likely to occur. Thus, HAeBe stars are 
natural candidates for studying the physical processes playing an important role 
in planet formation. Stars of spectral type A have gained a renewed interest because of the recent direct
imaging detection of extrasolar planets around these stars \citep{ref:marois2008}.

In this study, we focus on a subgroup of the HAeBe class with spectral type between late B and A-F, 
i.e., the lower mass end of the HAeBe class(hereafter HAe stars). Spectral energy distributions (SEDs) of these stars 
 can be well represented with models of a passive protoplanetary disk with a puffed-up inner 
rim \citep{ref:ddn2001}. Based on observations with the {\it Infrared Space Observatory} (ISO),
\citet{ref:meeus2001} classified the HAe stars into two groups. SEDs of {\it Group II} sources 
can be well fitted with a power law at mid- to far-infrared wavelengths. An additional blackbody 
component is required, however, to fit the SEDs of {\it Group I} sources at far-infrared wavelengths. 
Theoretical models of protoplanetary disks showed that SEDs of {\it Group I} sources can be explained
by flared disks, which are in vertical hydrostatic equilibrium and where gas and dust are well mixed. 
 Later on as dust grains grow in size and settle to the mid-plane, the disk becomes flatter producing
the steeper, bluer mid- to far-infrared SEDs of {\it Group II} sources \citep{ref:dullemond_dominik2004}.

The global shape of the SED, however, carries only limited information about the physical properties 
of protoplanetary dust grains and the processes they undergo. Mid-infrared spectroscopy, on the 
other hand, is an excellent diagnostic tool for studying the size, shape and chemical composition of 
protoplanetary dust grains. The mid-infrared domain is rich in vibrational resonances of silicates with 
different compositions, which are the main constituents of the protoplanetary dust (Henning 2009). 
These mid-infrared emission features originate in the hot surface layer of the disks where the temperature 
is above $\sim$100\,K. Since this region of the disk is optically thin, by analyzing the emergent spectra, the 
composition of the dust mixture as well as the physical parameters of the radiating dust grains (e.g., size or 
shape) can be derived. Mid-infrared spectroscopy, however, has also limitations. It is sensitive only to dust 
grains which show resonances in the mid-infrared, i.e., grains larger than several microns or "featureless"
grains (e.g., amorphous carbon or iron) cannot be studied in this way. Since mid-infrared features arise from the surface
layers of the disk, the derived grain properties are not necessary representative for the whole vertical 
extent of the disk.

Crystalline silicates are abundant in many solar system comets (see e.g., \citet{ref:wooden2007} and references therein)
but they are essentially missing from the interstellar medium (ISM). From the analysis of the 10\,{\micron} silicate
band \citet{ref:kemper2005} and \citet{ref:min2007} placed an upper limit of 2\,\% in terms of 
mass for the abundance of silicate crystals in the ISM. Although they represent usually a minor dust constituent 
in terms of abundance, the sharp features of crystalline silicates are frequently observed toward 
young stars, including HAe stars (e.g. \citet{ref:bouwman2001}, \citet{ref:van_boekel2005}). It is therefore
reasonable to assume that crystallization occurs in the disks of young stars. Due to their sharp features
crystalline silicates can be used as tracers to investigate dynamic processes in protoplanetary disks.
Both ways of crystal formation (annealing and direct condensation from the gas phase) require high temperature, 
typically above 1000\,K \citep{ref:fabian2000}. The fact that we still observe crystals in the outer disk where the temperature
is of the order of 100\,K suggests, that either a large-scale mixing should occur in the disks of young
stars \citep{ref:gail2004} or amorphous grains should be heated locally (e.g. shocks) to be transformed into crystals
(e.g. \citet{ref:harker_desch2002},  \citet{ref:sargent2009}).

Grain growth is another important process in protoplanetary disks which can be studied 
by mid-infrared spectroscopy. As sub-micron sized amorphous grains grow in size above a micron 
their 10\,{\micron} silicate feature becomes broader and flatter compared to the triangular shaped
feature of the smaller grain population. This was indeed observed in the spectrum of many young stars
regardless of their spectral type (e.g. \citet{ref:van_boekel2005, ref:apai2005, ref:bouwman2008, ref:watson2009}).
It was also reported by \citet{ref:bouwman2008} and \citet{ref:meeus2009} that the size of
the dust grains tends to be larger in flatter disks compared to flared one. This is the first observational
evidence that dust sedimentation can be the reason why initially flared disks evolve to flatter ones. 

In all of the above mentioned studies the average signal-to-noise ratio (S/N) of the mid-infrared spectra
was of the order of $\sim$100 or lower. In this paper, we take one step further and analyze Spitzer IRS spectra 
of a comprehensive set of HAe stars with extremely high quality (with S/N up to several hundreds). 
The goal of our analysis is to (1) test our knowledge collected from analysis of lower S/N data and (2) 
look for possible new dust species/effects, which are rare/weak enough to be observable only in high 
quality data, (3) investigate if the relationship between disk flaring and grain size, found by \citet{ref:bouwman2008},
\citet{ref:meeus2009} and  \citet{ref:sargent2009b} for T Tauri stars, also holds for HAe stars. 
 The analysis of the PAH emission will be presented in a separate paper (Acke et al., in prep).

\section{Observations}

\subsection{Sample selection}

The list of sources was compiled from the samples of \citet{ref:the1994}, \citet{ref:sylvester1996}, 
\citet{ref:van_den_ancker1998}, \citet{ref:sylvester_mannings2000} and \citet{ref:malfait1998}.
This source list was cross-correlated with the Spitzer Archive\footnote{Most of the observations came from
two programs, PI: J. Bouwman, PID:3470 and PI: B. Acke, PID:20308} and the observed sources were selected. 
Since the five studies, from which our original sample was derived, used different classification criteria 
HAeBe stars, the sample was not uniform and false classifications occurred. Sources which were in fact
not Herbig Ae stars (but e.g., classical Be systems or asymptotic giant branch stars) were rejected from our sample. 
Sources with obvious extended emission were also rejected, since our goal was to study the dust 
properties in \emph{disks} around HAe stars. The resulting final sample consists of 53 sources in 
total out of which 45 shows silicate emission features while the remaining eight sources show 
only emission lines from polycyclic aromatic hydrocarbons (PAHs).  Although in some "PAH-only"
sources there may be a hint of a weak 10\,{\micron} silicate feature the feature is so weak/shallow that a
meaningful dust composition cannot be determined we therefore excluded them from the final sample.
The final list of targets is presented in Table\,\ref{tab:source_list}.  In this paper we focus on the sources 
showing silicate emission and the analysis of the PAH bands will be presented in a forthcoming paper (Acke et al., in prep).

\subsection{Data reduction}
\label{sec:data_reduction}

The spectra presented in this paper were obtained using the {\it Infrared Spectrograph} \citep[IRS][]{ref:houck2004} 
on board the Spitzer Space Telescope. In most cases observations were performed using the short-low module (5.2--14.5\,{\micron}) 
of the low-resolution (R=$\sim$60--120) spectrograph and both the short-high (9.9--19.5\,{\micron}) 
and long-high (18.7--37.2\,{\micron}) modules of the high resolution (R=600) spectrograph. In the case of HD152404
only low-resolution modules, short-low and long-low (14-35\,{\micron}), were used. For 8 sources there were no low-resolution spectra taken 
with the Spitzer IRS instrument, only the short-high and the long-high modules were used.

In the case of low-resolution mode the data reduction process started from the {\tt droopres} intermediate data product processed
through the SSC pipeline S15.3.0. Our data are further processed using spectral extraction tools developed for the
FEPS {\it Spitzer} science legacy program, partially based on the {\tt SMART} software package \citep{ref:higdon2004}.
Most of our observations were taken in standard staring mode where the target is observed
at the two nominal nod positions in the slit ($\sim$18" from the slit center), using multiple cycles per target for redundancy and 
to allow the rejection of artifacts introduced by bad pixels or cosmic ray hits. A high accuracy IRS or PCRS
peak-up (with a 1$\sigma$ pointing uncertainty of 0.4" radius) was used to acquire targets in the spectrograph slit. 
A subset of our sources has been observed in 2$\times$3 mapping mode without a peak-up. The small maps consist of two
positions at the nominal nod positions in slit, similar to normal staring observations, and three map positions in a 
perpendicular direction to the slit, with the central position centered on the target and the other positions shifted by 
half a slit width (1.8"). Effectively, this results in three standard staring mode observations with one observation 
reasonably centered on the source and two offset observations. We have used the central map position and use those as normal 
nodded observations in standard staring mode.

As a first step, we correct for the background emission and stray light (mainly coming from the infrared background seen 
by the peak-up array) by subtracting the associated pairs of imaged spectra of the two nodded positions along the slit for 
each module and order. Pixels flagged by the data pipeline as being "bad" were replaced with the average pixel value of a six 
pixel elongated box surrounding the bad pixel. The method we apply for finding the mean pixel value resembles Nagao \& Matsuyama 
filtering \citep{ref:nagao1979} and ensured edge preservation in the source region of our spectral images. The spectra were extracted 
using a fixed-width aperture of six pixels centered on the position of the source. The exact source position relative to the slit 
was determined by fitting a sinc profile to the spectra in the dispersion direction using the collapsed and normalized source profile.

The spectra are calibrated with a relative spectral response function derived from IRS spectra and MARCS stellar models for a 
suite of calibrators provided by the Spitzer Science Center through the Spitzer data archive. The spectra of the calibration stars 
($\eta$1~Dor, HR~6606, HR~7341) were extracted using the same method as for our science targets. One of the most 
difficult problems with spectroscopy using a narrow slit is the spectro-photometric calibration. Due to telescope pointing 
uncertainties and drifts, a variable fraction of source flux is being blocked by the slit. For high accuracy peak-up observations 
the intrinsic photometric accuracy is about 10\%, while observations with no peak-up have a far lower accuracy. Due to the 
wavelength dependence of the point spread function (PSF) these pointing-induced flux losses will also change the spectral shape. 
To remove any effect of pointing offsets, we developed a correction method based on the PSF of the IRS instrument, 
correcting for possible flux losses. For details of this method we refer to \citet{ref:swain2008}.
We estimate the flux accuracy we can achieve with our data using this method to be 1\%.  

The data reduction procedure for the high-resolution data was based on the method developed by the Cores-to-Disks Spitzer 
legacy team \citep{ref:lahuis2007}. The procedure started from the {\tt rsc} products processed through the same version (S15.3.0) of 
the {\it Spitzer} data pipeline as the low-resolution data. The spectra were extracted in two ways. The first method uses a fixed 
width aperture very similar to the method we used for the low-resolution data. The second method is an optimal source profile 
extraction method which fits an analytical PSF derived from sky-corrected calibrator data and an extended emission 
component, derived from the cross-dispersion profiles of the flat-field images, to the cross-dispersed source profile. It is not
possible to correct for the sky contribution in the high-resolution spectra, subtracting the two nod positions as with the 
low-resolution observation, due to the small slit length. We either subtracted an observation on the sky at a position close to 
the source or, when no such sky observation was taken, used the background estimate from the source profile fitting extraction 
method. For correcting "bad" pixels we used the IRSCLEAN package. We further removed low-level ($\sim$1\%) fringing using the {\tt irsfringe} 
package \citep{ref:lahuis_boogert2003}. We carefully checked that our fringe removal was not affecting the multiple silicate bands 
seen in our spectra. As the frequency of the fringes is reasonably well constrained and higher than the typical width of the 
observed thermal emission features from the various dust components, we found this not to be a problem.

The flux calibration for the high-resolution spectrograph has been done in a similar way as for the low-resolution observations. 
For the relative spectral response function we also used MARCS stellar models and calibrator stars provided through the Spitzer 
Science Center. The spectra of the calibration stars were extracted in an identical way to our science observations using both 
extraction methods. As with the low-resolution observations, we also corrected for possible flux losses due to pointing offsets. 
We estimate the absolute flux calibration uncertainty for the high resolution spectra to be $\sim$3\%, slightly higher than that 
of the low-resolution observations. We found that the fixed width aperture extraction gave the best result for the short-high 
module (9.9--19.5\,{\micron}), while for the long-high module (18.7--37.2\,{\micron}) the optimal extraction method was slightly 
better. For the final spectra presented in this paper we therefore used the results of the fixed width aperture extraction for 
the wavelengths shortward of 19\,{\micron}, and the results of the optimal extraction method for wavelengths longward of 
19\,{\micron}. We want to note that in a few spectra (HD35187, HD38120, HD139614) the spectrum in the 12th order ($\sim$31--34\,$\mu$m) 
of the long-high module seems to be tilted compared to the neighboring orders using the optimal extraction method. 
In the case of HD139614 we saw similar behaviour in the 15th order of the long-high module ($\sim$25--27\,$\mu$m).
Although the full aperture extraction method did not show such strong tilt, it gave significantly higher noise level in this order, 
than the optimal extraction. Since the choice of the extraction method did not change our results we used the optimal extraction method 
for the long-high module to obtain uniformly reduced data in the whole sample. 

After the spectra have been reduced the different modules were combined to achieve our final spectra. Between 5.5 and 13.5\,{\micron} 
the short-low module was used while we used the short-high and the long-high for the 13.5--19.5{\micron} and 19.5--35\,{micron} 
wavelength intervals, respectively. For the sources, where no low-resolution Spitzer IRS spectra were taken, 
Spitzer spectra were supplemented shortward of 13.0\,{\micron} by data taken with the TIMMI2 instrument 
from \citet{ref:van_boekel2005}, if such data were available. 
The high-resolution and the TIMMI2 spectra were rebinned for a uniform spectral resolution of R=160 
for the spectral fitting. Though the absolute flux calibration of the IRS observations is very good, any differences in the absolute 
flux calibration in various modules were handled in the following way. The spectra in different modules were scaled to a reference
module which is chosen to be the one with the lowest absolute flux calibration uncertainty. We used, therefore, the short-low module
as a reference, if it was present. If no short-low module was available the short-high module was chosen to be the reference. The applied
scaling factors are of the order 1.

\section{Analysis}
\subsection{Dust model}

In order to study evolution and thermal processing of protoplanetary dust grains using mid-infrared 
spectroscopy, first one needs to identify the abundant dust species in the disks around young stars.
Such an identification can be done by comparing the laboratory measurements of mass absorption 
coefficients (MACs) of different materials to the emission features observed in the spectra. Such a comparison/identification
has already been done by e.g., \citet{ref:molster2002}. These studies showed that mid-infrared spectra of
young stars can be well reproduced by a mixture of five dust species, amorphous silicates with olivine 
and pyroxene stoichiometry, crystalline forsterite and enstatite and silica.  The IRS instrument on board  
the Spitzer Space Telescope allowed us to improve the S/N of mid-infrared spectra 
by more than an order of magnitude compared to ISO SWS and and by a factor of 5--8 compared to 
ground-based instruments (e.g., COMICS, TIMMI2, T-ReCS). The exercise was repeated on the Spitzer data
and emission features seen in the spectra were identified. The identification of the features is summarized 
in Tab\,\ref{tab:feature_list}. The dust features seen in our spectra
can be identified as any of the following materials: amorphous silicates with olivine and pyroxene stoichiometry, 
forsterite, enstatite, silica and PAHs.

Amorphous silicates of olivine (Mg$_{\rm x}$Fe$_{\rm1-x}$SiO$_4$) and pyroxene (Mg$_{\rm x}$Fe$_{\rm1-x}$Si$_2$O$_6$) type are 
represent more than about 98\% of silicate dust  grains\footnote{Here, we neglected all "featureless" dust species 
(e.g., iron and carbon) for which only weak constraints can be drawn from mid-infrared spectroscopy} in the ISM 
\citep{ref:kemper2005, ref:min2007}, where the 
protoplanetary dust grains are thought to originate. Olivine-type amorphous silicates show a broad triangular-shaped
feature in the 10\,{\micron} region which peaks at 9.8\,{\micron}. Pyroxenes show a similar band to olivine 
in the 10\,{\micron} region, but its peak position is located at somewhat shorter wavelengths 
($\sim$9.2\,{\micron}). 

The broad features of amorphous silicates are less sensitive to the applied scattering theory (grain shape effects) than 
crystalline bands. It is thus not surprising that, apart from the size of the grains, not much information is available on 
the properties (e.g., shape, Mg-content) of the amorphous silicate grains.  For instance, most of the studies 
(e.g., \citet{ref:van_boekel2005}, 
\citet{ref:bouwman2008}) used the optical constants of iron-magnesium silicates with Fe/(Mg+Fe)=0.5 published by 
\citet{ref:dorschner1995} with Mie theory, assuming compact spheres for the grain shape. 
The aforementioned iron content of the silicates was used on the basis of cosmic element abundance constraints, 
and their higher mid-infrared opacities, compared to iron-free silicate grains.
It is, however, surprising that although protoplanetary dust grains are always regarded as porous aggregates  
(e.g., \citet{ref:henning_stognienko1996})
a compact sphere model can fit the observed features relatively well. From the analysis of the 10\,{\micron} silicate 
absorption profile 
toward the Galactic Center \citep{ref:min2007} concluded that the best fit can be obtained by using porous iron-free silicates.
We used both iron-magnesium silicates with Fe/(Mg+Fe)=0.5 and iron-free silicates with Fe/(Mg+Fe)=0 and systematically tested
the Mg-content of the amorphous grains and the scattering theory (i.e. grain shape effect).

In contrast to the broad features of amorphous silicates, crystalline silicates show sharp and narrow features 
in the mid-infrared, which can be frequently seen in the spectra of both young and evolved stars \citep{ref:henning2009}. 
The analysis of the positions and the relative strength of these sharp features revealed that the radiating material 
should be a mixture of forsterite (Mg$_2$SiO$_4$) and enstatite  (MgSiO$_3$, see e.g., \citet{ref:malfait1998b, ref:bouwman2001, 
ref:meeus2001}).
These minerals are the magnesium-end members of the olivine and pyroxene solution series. Although crystalline 
silicates are usually minor dust components in protoplanetary disks compared to amorphous silicates, their
sharp features can be seen in the spectrum in almost all cases. 

Studies of interplanetary dust particles show that these grains frequently contain large inclusions of silica.
Laboratory annealing experiments of amorphous silicates also show that during the formation 
of forsterite, silica can be produced (e.g., \citet{ref:fabian2000}). Indeed silica has been found in the spectrum of young stars 
both 
in amorphous and in crystalline form (e.g., \citet{ref:van_boekel2005} or \citet{ref:sargent2009}).  
Silica shows a narrow, strong distinct features at $\sim$9\,{\micron} and a broad, but also strong band at $\sim$21\,{\micron}.  
The dust species together with the references of the applied optical constants are summarized in Table\,\ref{tab:optconst}.
Apart from the above-mentioned five dust species, we did not find any evidence for other abundant dust species in 
the Spitzer data. 

\begin{figure} [!ht]
\includegraphics[scale=0.6, angle=0]{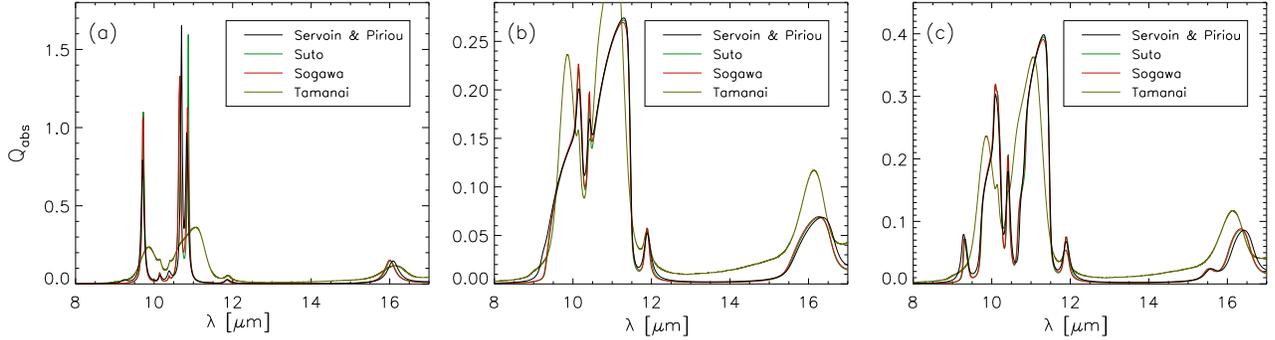}
\caption{Comparison of scattering theories and different sets of optical constants for 
crystalline forsterite (\citet{ref:servoin1973, ref:sogawa2006, ref:suto2006}). 
The applied scattering theories were, ({\it a}) Mie theory, ({\it b})
CDE, and ({\it c}) DHS with a maximum volume filling factor of 1.0. For comparison the
optical efficiencies of forsterite measured on free-flying particles \citep{ref:tamanai2009} are shown.}
\label{fig:opac_forsterite_short}
\end{figure}

\begin{figure} [!ht]
\includegraphics[scale=0.6, angle=0]{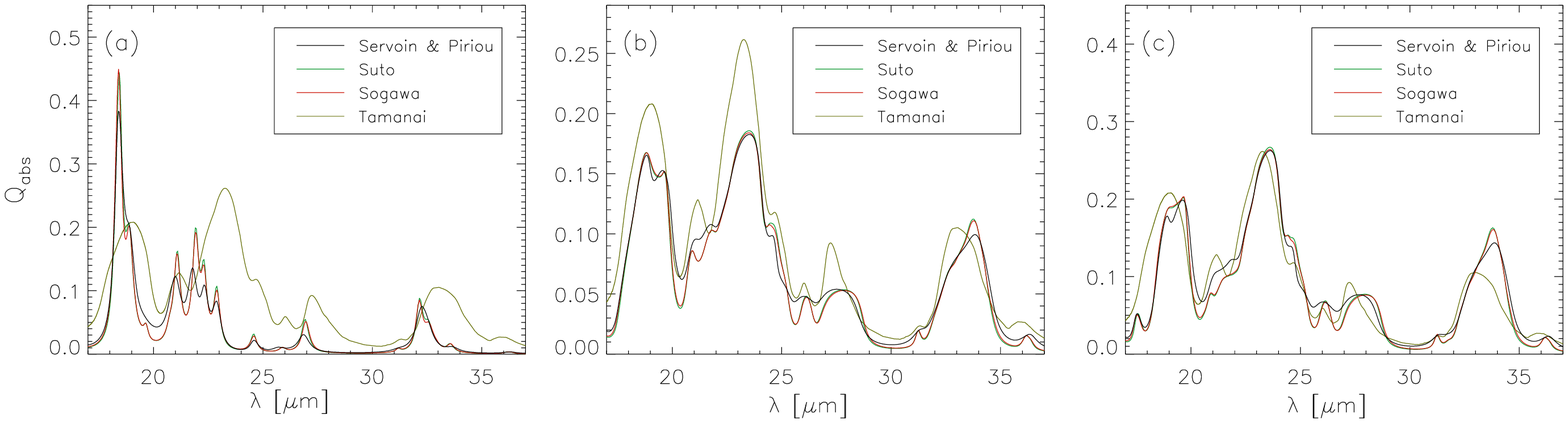}
\caption{Same as Figure\,\ref{fig:opac_forsterite_short}, but for longer wavelengths.}
\label{fig:opac_forsterite_long}
\end{figure}

Three scattering theories were considered to calculate  (MACs) from the optical constants, 
Mie scattering, continuous distribution of ellipsoids (CDE) and distribution of hollow spheres (DHS). 
These scattering theories are the most widely used methods to model mid-infrared spectra of young stars.
We have two requirements for the computation method we wish to apply during the analysis. 
(1) The shapes and positions of the dust features in the Spitzer spectra should be reproduced as well as possible 
(2) The applied theory should also be valid outside of the Rayleigh limit. 
The reason for this second requirement is that we wanted to study the sizes of dust grains. In the strong bands
at 10\,{\micron} one can already be outside of the Rayleigh limit for a micron-sized particle.
The comparison of band position found in the spectra and those in calculated MACs rules out the Mie theory immediately
(see, Table\,\ref{tab:feature_list}). In Figures\,\ref{fig:opac_forsterite_short} and \ref{fig:opac_forsterite_long} 
we present the calculated absorption efficiencies of forsterite using different scattering theories and 
compare them to laboratory measurements from \citet{ref:tamanai2009}.
It can be seen, that dust band positions can be about as well matched with DHS as with CDE, since the calculated MACs 
do not differ so much from each other than they do compared to Mie scattering  (see also \citealt{ref:min2003}). 
Our second requirement, however, excludes 
CDE since it is strictly valid within the Rayleigh limit only. In the case of DHS both of our required conditions are fulfilled, and
furthermore it is a fast computational method. We used, therefore, the DHS theory to calculate the MACs
from the optical constants for our analysis. MACs  of each dust species were calculated for three discrete grain sizes 
(0.1\,{\micron}, 2.0\,{\micron} and 5.0\,{\micron}). For forsterite and enstatite we used only two grain sizes (0.1\,{\micron} and 
2.0\,{\micron}) as we did not find any evidence for large ($>2.0$\,{\micron}) crystals. Silicate grains larger than $\sim$5\,{\micron} are not 
considered, since they do not show feature in the studied wavelength range (5--35\,{\micron}). 

\begin{figure}[!ht]
\includegraphics[scale=0.56, angle=0]{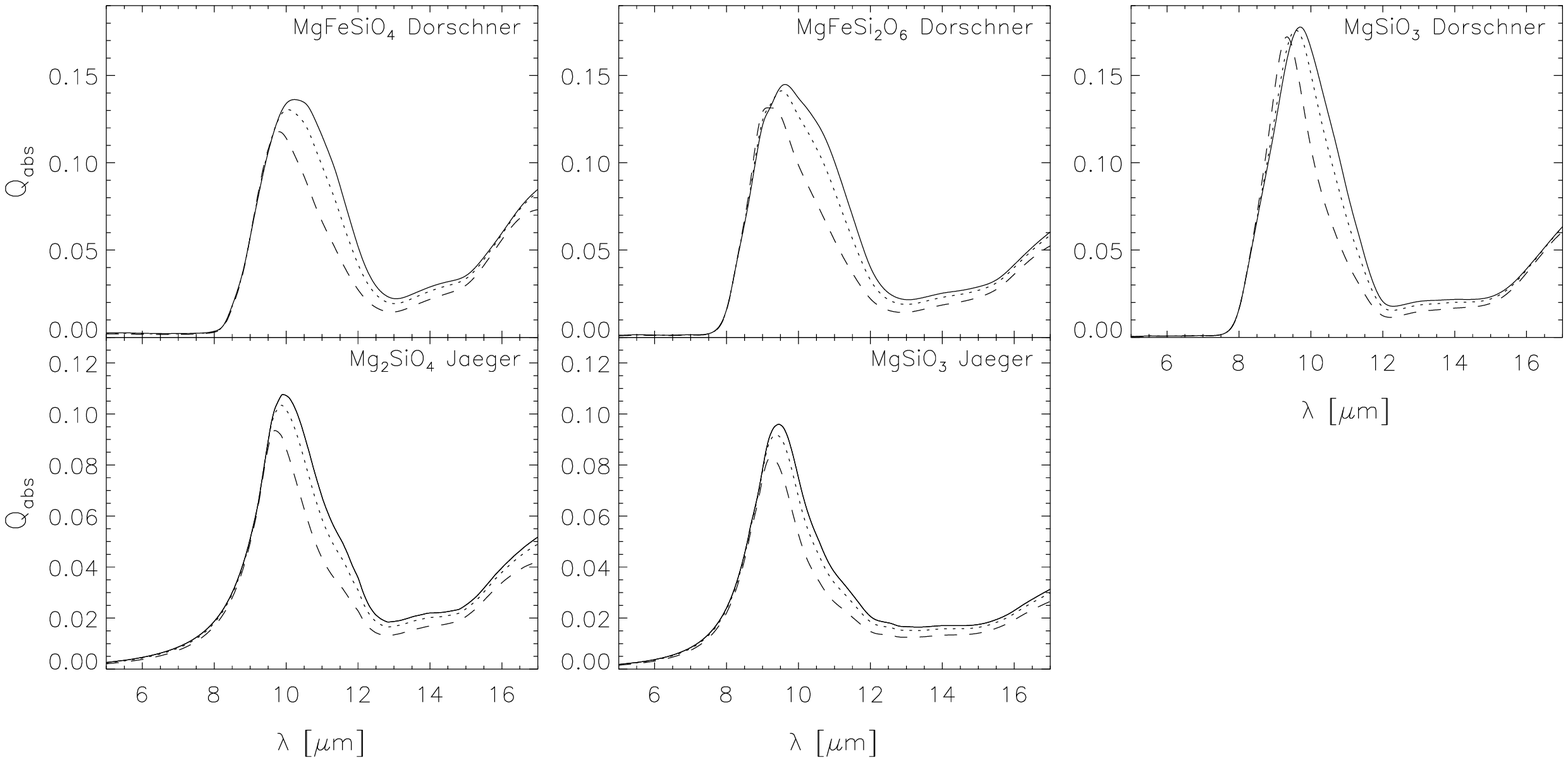}
\caption{Absorption efficiencies of the amorphous silicates. Solid, dotted and dashed lines
show the absorption efficiencies calculated from the optical constants (see Table\,\ref{tab:optconst}) 
using the DHS theory with a maximum volume filling factor of 1.0, 0.7 and 0.0, respectively. 
The zero filling factor solution is identical to the solution of Mie theory.}
\label{fig:opac_amorph}
\end{figure}

In DHS one computes the scattering/absorption cross section of hollow spheres with a volume fraction $f=V_{\rm tot}/V_{\rm vac}$, 
where $V_{\rm tot}$ is the total volume of the grain and $V_{\rm vac}$ is the volume of the vacuum inclusion. 
The final MACs will then be an average over a whole distribution of hollow spheres with different values of $f$. 
It has already be shown that for crystalline silicates one should average over all possible values of $f$ (from 0
to 1.0) to get the best agreement with the observed positions of crystalline bands (see e.g.,, \citet{ref:min2003}). 
In Fig\,\ref{fig:opac_amorph} we show the absorption efficiencies of amorphous silicates with olivine and pyroxene
stoichiometry calculated using DHS theory. It can be seen that the higher the upper boundary for the hollow
sphere distribution ($f_{\rm max}$) is chosen the broader the feature becomes. By increasing the value of 
$f_{\rm max}$, the peak position of the feature shifts toward longer wavelengths. 
For the amorphous silicates we found that the best agreement with the observed spectra is obtained if
one uses $f_{\rm max}=0.7$. For the details, see Sec\,\ref{sec:amorphous_silicates}.

\subsection{PAH band profiles}
\label{sec:pah_profiles}
\begin{figure} [!ht]
\includegraphics[scale=0.8]{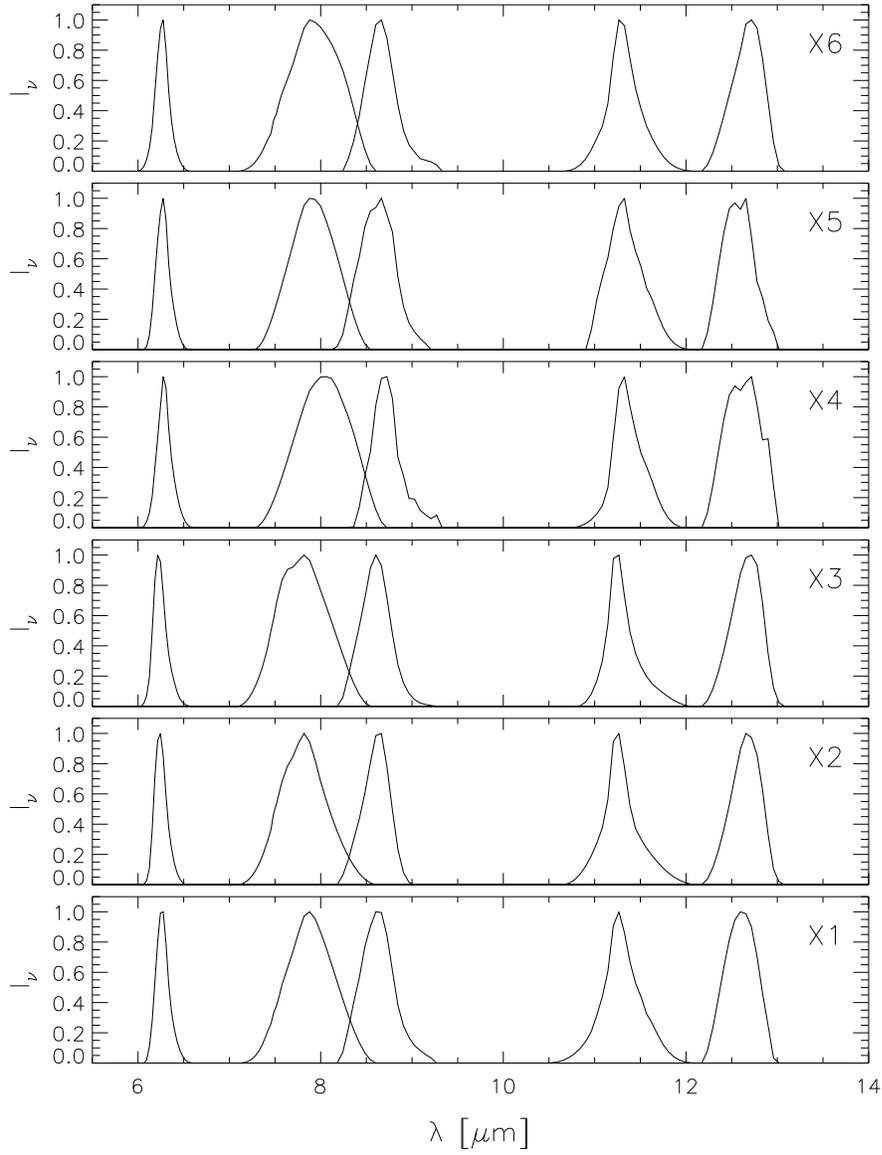}
\caption{PAH emission profiles used for the analysis. All band profiles are normalized
to their peak-value. For the details of the derivation of the different profiles and the notation of the
profile names (X1--X6) see Section\,\ref{sec:pah_profiles}}.
\label{fig:pah_profiles}
\end{figure}

All sources discussed in this paper show emission from PAHs.
PAHs are also included in the spectral decomposition procedure in order to avoid systematic biases in the estimated
dust parameters due to the PAH emission. PAH emission at 11.3\,{\micron}, 8.6\,{\micron} and 12.7\,{\micron} can cause 
confusion in the estimated forsterite and silica content, respectively.  In order to get the most realistic intensity profile
for the observed PAH features, band profiles have been extracted from the spectra of sources with PAH emission only.
These sources were HD34282, RR Tau, HD97048, HD135344B, HD141569 and HD169142.  Five band profiles have
been derived from the spectra of each source separately. We denote a set of profiles belonging to one source 
X1...X6, corresponding to HD34282....HD169142, respectively and we call the individual profiles after the central 
wavelength position as 6.2\,{\micron} 7.7\,{\micron}, 8.6\,{\micron}, 11.3\,{\micron} and 12.7\,{\micron} 
profiles (see Figure\,\ref{fig:pah_profiles}). The X1-6.2\,{\micron} profile is therefore derived from HD34282 and its central wavelength is about 6.2\,{\micron}. 
For further details of the derivation of the band profiles we refer to  Acke et al. (in prep).

\subsection{Spectral analysis}
In order to analyze the dust composition in the disk atmosphere, the radiation of which dominates the IRS
spectrum, we used the two-layer temperature distribution (TLTD) method described in \citet{ref:juhasz2009}. This method
uses a multi-component continuum (star, inner rim, disk midplane) and it assumes that the region where the 
observed radiation originates (both optically thin and thick) has a distribution of temperatures instead of a single
one.  In this fitting method the observed flux density at a given frequency is given by 

\begin{eqnarray}
\nonumber F_\nu = F_{\nu, {\rm cont}} & + & \sum_{i=1}^N\sum_{j=1}^MD_{i,j}\kappa_{i,j}
\int_{\rm{T_{\rm a, max}}}^{\rm{T_{\rm a, min}}}\frac{2\pi}{d^2}B_\nu(T){T}^{\frac{2-qa}{qa}}dT\\
 &+& \sum_{i=1}^{NP}C_{i}I^{\rm PAH}_{i}
\label{eq:1}
\end{eqnarray}

where, $N$ and $M$ are the number of dust species and grain sizes, respectively.
$N_{p}$ denotes the number of different PAH templates included in the fit. 
$\kappa_{i,j}$ is the mass absorption coefficient of the dust species $i$ and grain size $j$. $B_\nu(T)$
is the Planck function, $qa$ is the power exponent of the temperature distribution and $d$ is the distance
to the source. The subscript $a$ in the integration boundaries refers to the disk atmosphere. The continuum 
emission ($F_{\nu, {\rm cont}}$) is given by

\begin{eqnarray}
\nonumber F_{\nu, {\rm cont}} = \frac{\pi R_\star^2}{d^2} B_\nu(T_\star)&+& D1\int_{\rm{T_{\rm r,max}}}^{\rm{T_{\rm r, min}}}\frac{2\pi}{d^2}B_\nu(T){T}^{\frac{2-qr}{qr}}dT \\
&+& D2\int_{\rm{T_{\rm m,max}}}^{\rm{T_{\rm m, min}}}\frac{2\pi}{d^2}B_\nu(T){T}^{\frac{2-qm}{qm}}dT.
\label{eq:2}
\end{eqnarray}

The first term on the right hand side describes the emission of the star, while the second and third terms describe 
the radiation of the inner rim and the disk midplane, respectively. The stellar emission, used for the fits, was not fitted
during the mid-infrared spectral analysis, but it was derived from a separate fit to the UV-optical photometry from the literature.  

The assumptions (e.g., one single dust composition) used in the TLTD method are not valid for an arbitrarily
broad wavelength interval (see \citet{ref:juhasz2009}). Fitting the Spitzer IRS spectra to the total available wavelength
interval (5.5--35\,{\micron}) is already not reasonable. Therefore, we divided the Spitzer IRS wavelength range into 
two regions, 5.5--17\,{\micron} and 17--37\,{\micron}. These two wavelength
intervals were fitted separately, although for the longer wavelengths we used the star and the rim emission which were fitted
to the 5.5--17\,{\micron} range. PAH templates were included in the fit only for the shorter wavelength interval.
The final model in the 5.5--17\,{\micron} region was obtained using seven fits of each spectra. 
In the first fit only the X1 set of PAH band profiles was used, in the second fit we used only the X2 profiles, etc. 
After the spectra were fitted with all six sets of PAH band profiles separately, we calculated the $\chi^2$ of the fit for 
the wavelength interval of the individual PAH bands. In the seventh fit we used a combination of PAH profiles taking
the best-fit profile (with the lowest $\chi^2$) for each band 
(e.g., X1-6.2\,{\micron}, X5-7.7\,{\micron}, X6-8.6\,{\micron}, etc.). The final model for a given spectrum was chosen to
be the one which gives the lowest global $\chi^2$ for the whole 5--17\,{\micron} interval.
For the fits in the longer wavelength interval the rim contribution was not fitted, only the optically thin emission and the 
midplane component. We used the parameters for the rim which were derived from the fitting of the 10\,{\micron} region. 

 To estimate the uncertainties on the derived dust parameters we used a Monte Carlo type of error estimation
(e.g.,,\ \citealt{ref:van_boekel2005, ref:min2007}). In this kind of error estimation, a normally distributed noise is added to 
the spectrum, scaling the width of the distribution to the simulated observational uncertainty in the flux value.
Then the resulting spectrum is fitted. This procedure was repeated 100 times. Then the standard deviation of the resulting mass 
fractions from the 100 fits will be the uncertainty of the derived dust compositions.

\section{Results}
\subsection{General summary of the fits}

\begin{figure}
\includegraphics[scale=.75]{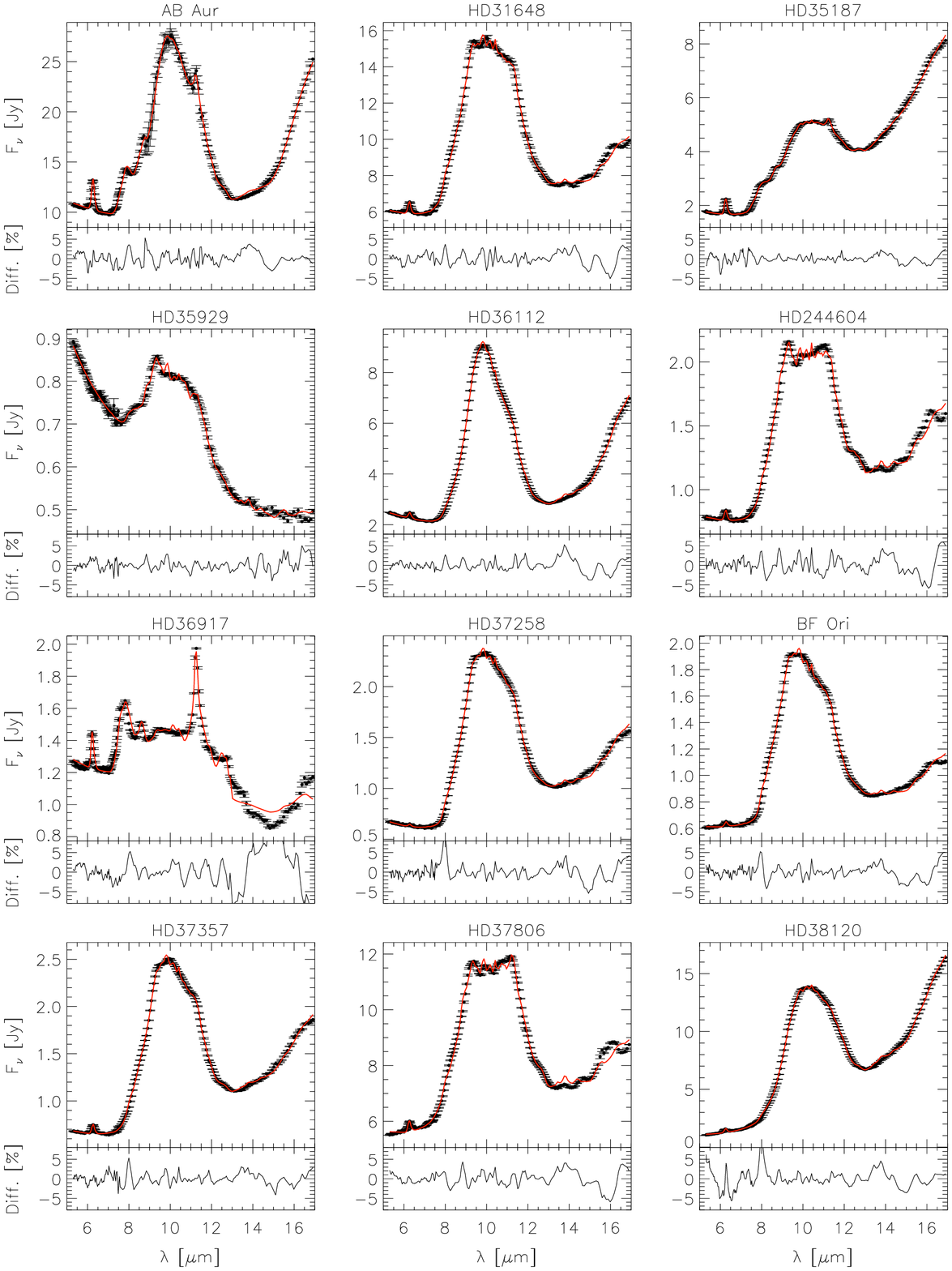}
\caption{Fits to the short wavelength range (5--17\,{\micron}). The observed
IRS spectrum is shown with black dots, while the red line shows the best-fit
model. For each fit the residuals ((F$_\nu^{\rm model}$-F$_\nu^{\rm obs}$)/F$_\nu^{\rm obs} \times 100$) are also shown. }
\label{fig:fit1}
\end{figure}

\begin{figure}
\includegraphics[scale=.75]{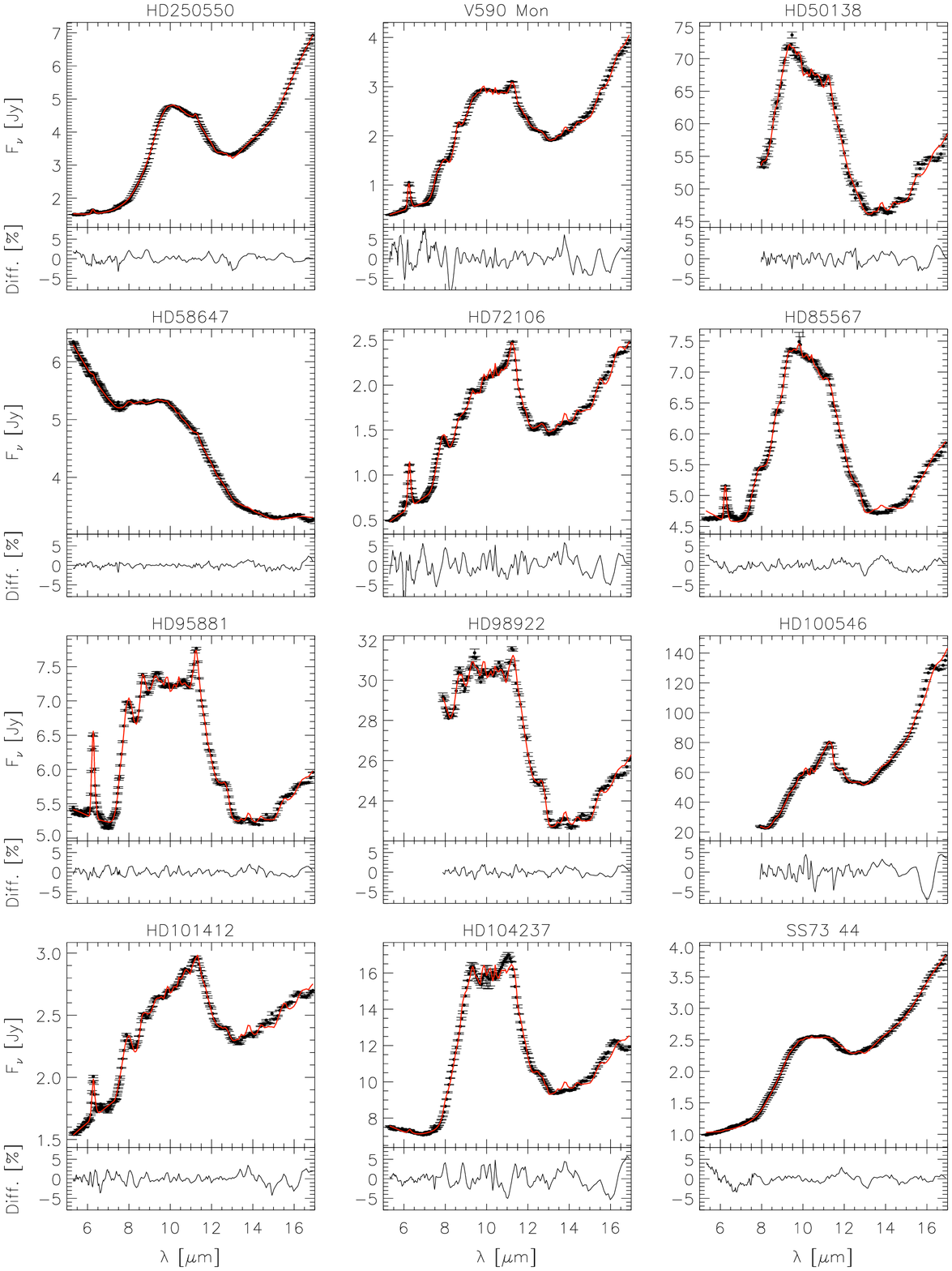}
\caption{Same as Figure\,\ref{fig:fit1}. }
\label{fig:fit2}
\end{figure}

\begin{figure}
\includegraphics[scale=.75]{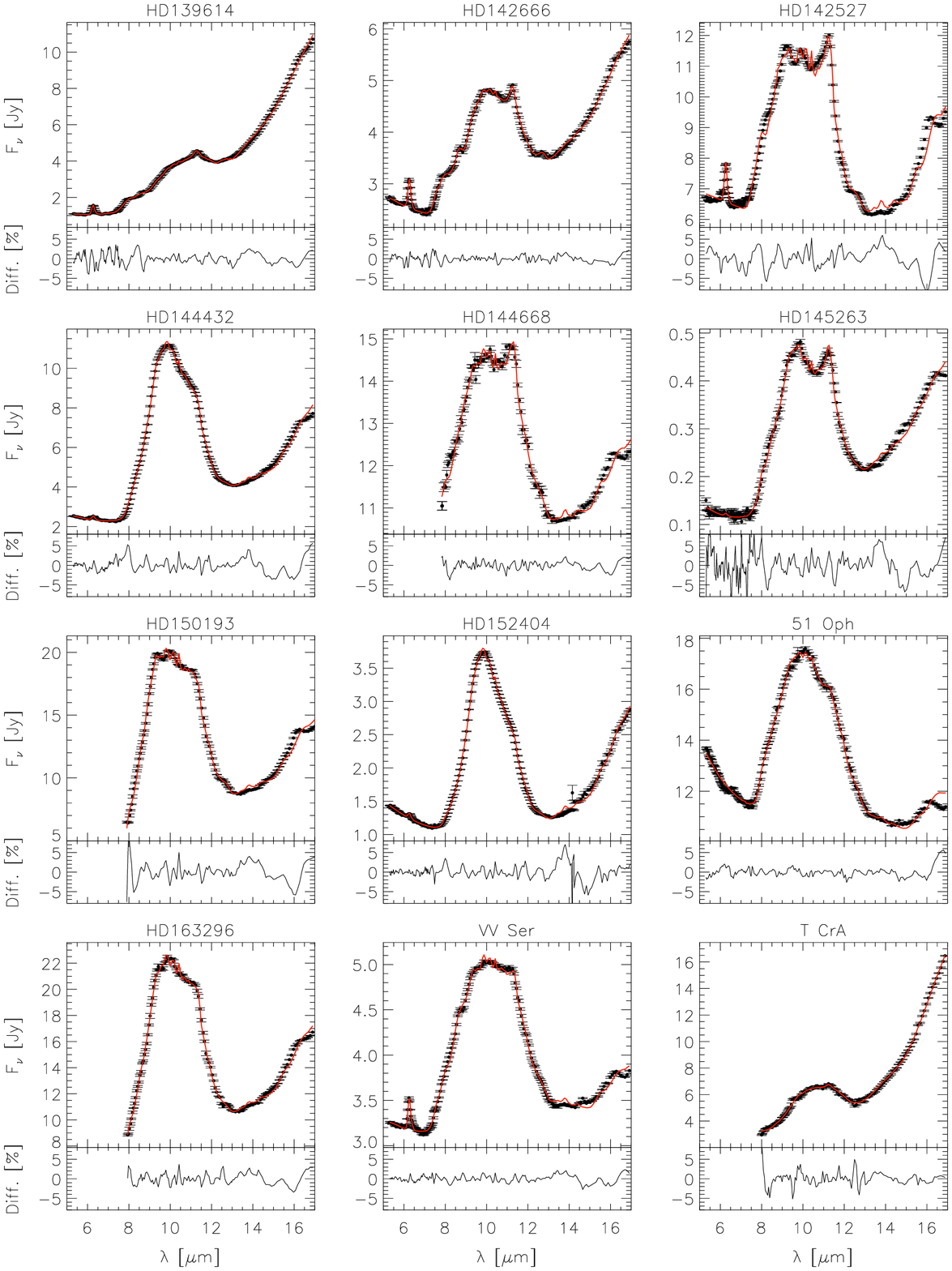}
\caption{Same as Figure\,\ref{fig:fit1}. }
\label{fig:fit3}
\end{figure}

\begin{figure}
\includegraphics[scale=.75]{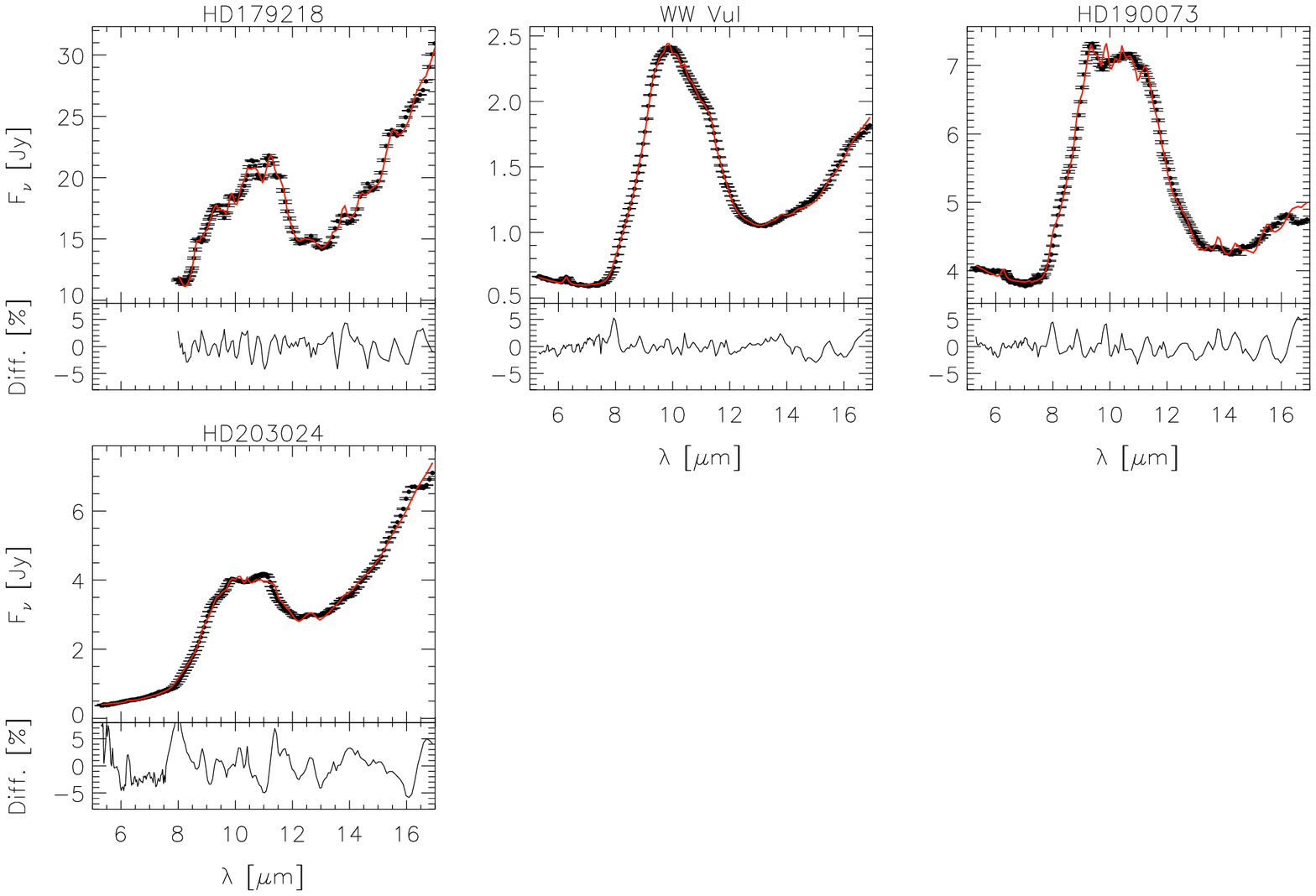}
\caption{Same as Figure\,\ref{fig:fit1}. }
\label{fig:fit4}
\end{figure}

\begin{figure}
\includegraphics[scale=.75]{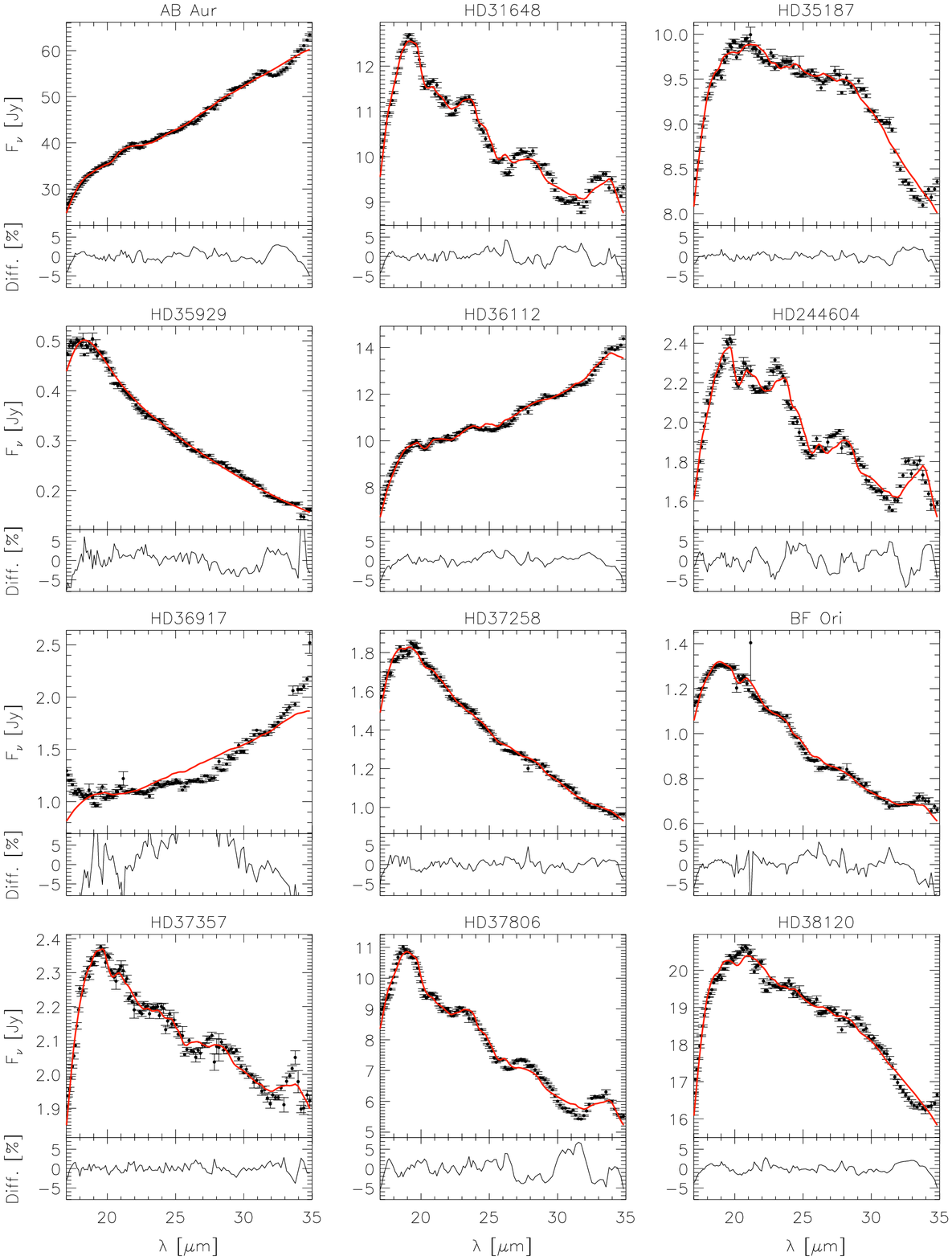}
\caption{Fits to the long wavelength range (17--35\,{\micron}). The observed
IRS spectrum is shown with black dots, while the red line shows the best fit
model. For each fit the residuals ((F$_\nu^{\rm model}$-F$_\nu^{\rm obs}$)/F$_\nu^{\rm obs} \times 100$) are also shown. }
\label{fig:fitl1}
\end{figure}

\begin{figure}
\includegraphics[scale=.75]{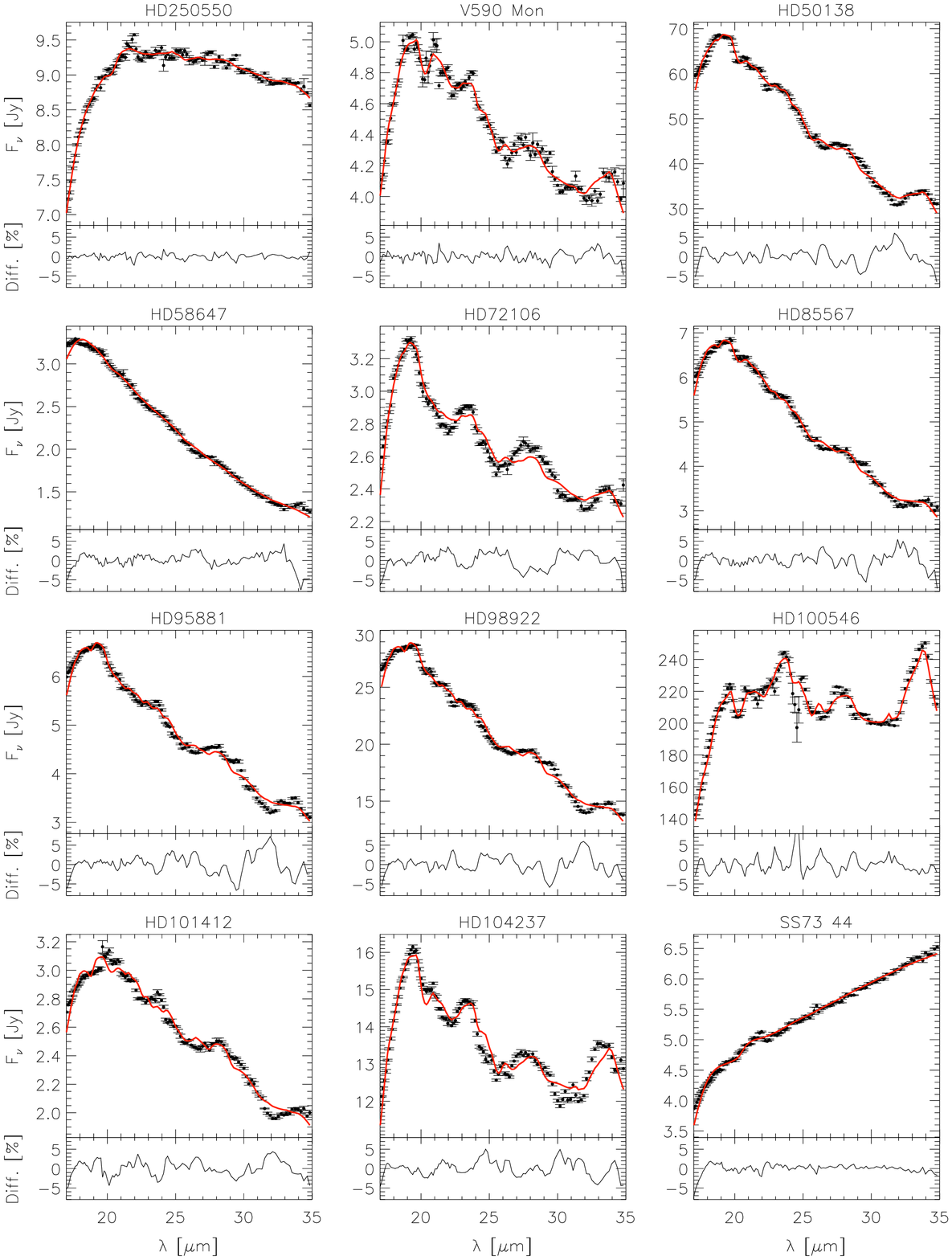}
\caption{Same as Figure\,\ref{fig:fitl1}. }
\label{fig:fitl2}
\end{figure}

\begin{figure}
\includegraphics[scale=.75]{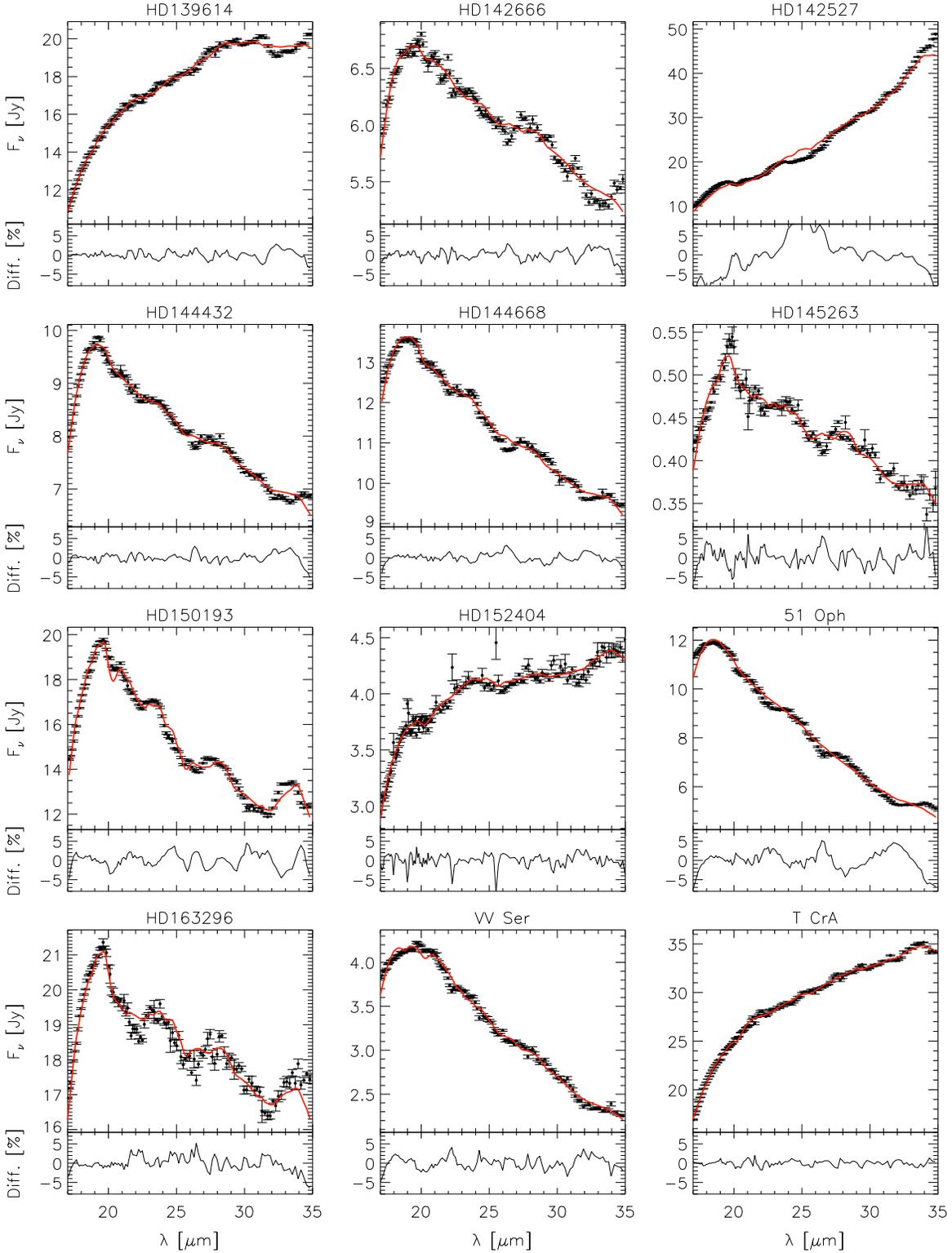}
\caption{Same as Figure\,\ref{fig:fitl1}. 
In the spectrum of HD139614 the emission feature between 26 and 31\,{\micron} is apparent and caused by problems
in the 15th and 12th order of the long-high module (see Section\,\ref{sec:data_reduction}).}
\label{fig:fitl3}
\end{figure}

\begin{figure}
\includegraphics[scale=.75]{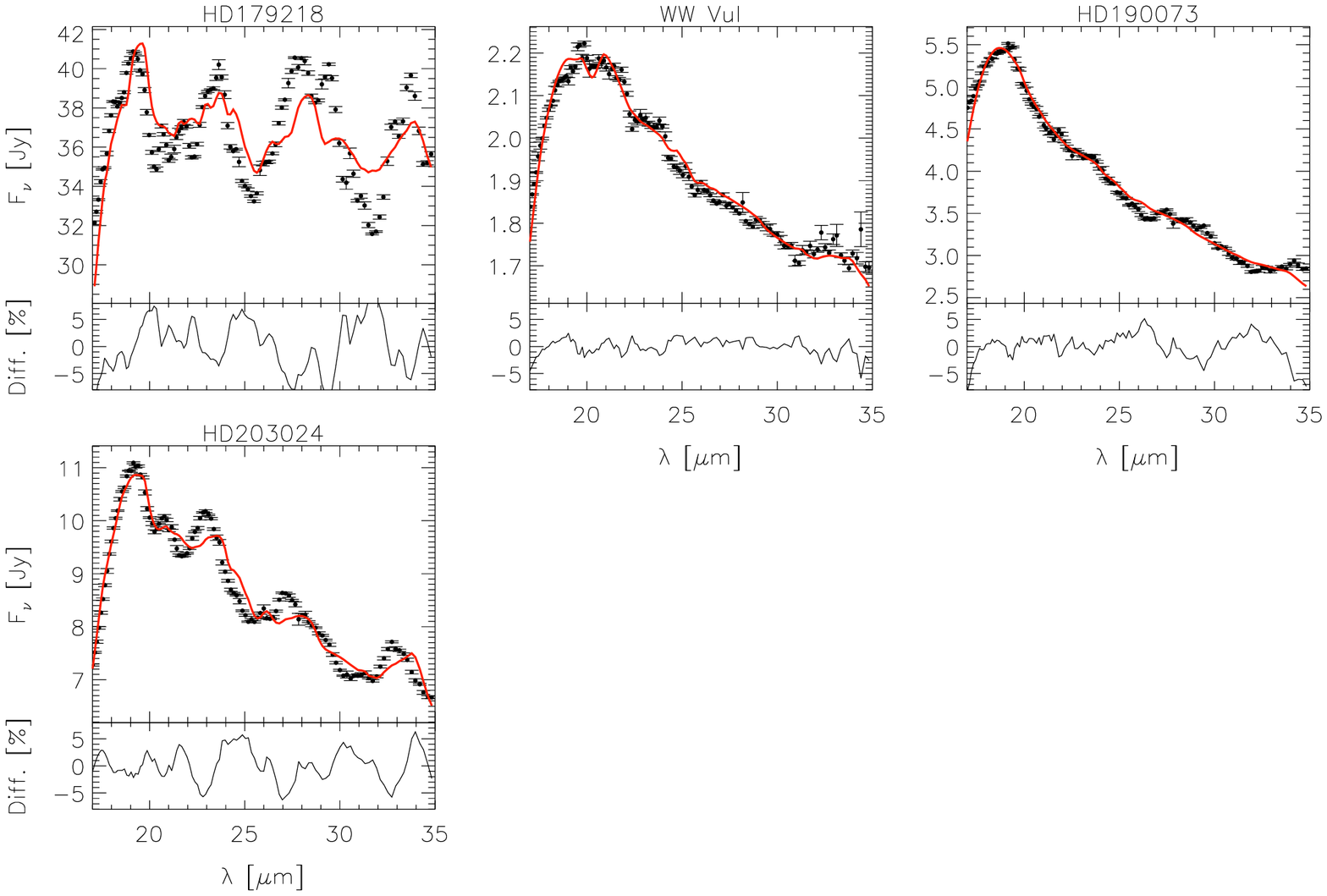}
\caption{Same as Figure\,\ref{fig:fitl1}. }
\label{fig:fitl4}
\end{figure}

The fitted dust composition for each source is presented in Table\,\ref{tab:res1}-\ref{tab:long_res6} while the fits themselves are shown 
in Figure\,\ref{fig:fit1}-\ref{fig:fitl4}. 
The agreement between the observed spectra and our models are very good in general, with only a few exceptions. 
 In three cases (HD35187, HD38120 and HD139614) a significant part of the $\chi^2$ in the long wavelength
fits originates in the region between 30\,{\micron} and 35\,{\micron} which is related to the problem with the 12th order of the long-high module 
(see Section\,\ref{sec:data_reduction}). In the case of HD36917 our model has difficulties to match the observed spectrum longward of 14\,{\micron}, 
which could be caused by the presence of a 16--19\,{\micron} PAH-band complex, which we did not take into account during the fitting. We believe, 
however, that these problems did not affect the main results of this paper. The reason is that the crystalline emission features in the long wavelength
interval, which are investigated in details later on, are either very weak or completely missing in the spectra of these sources. 

For the rest of the sample differences between model and observation 
are usually at the percentage level shortward of 17\,{\micron} and 5\,\%--8\,\% longward of 17\,{\micron}.  The reduced $\chi^2$ values
are, however, usually several tens in contrast to the expected value of about one for a good fit. We should keep in mind that the 
spectra analyzed in this paper have extremely high S/N (typically several hundreds). There are several effects 
which are negligible for lower S/N ratio spectra but become important for such extremely high S/N. 
The most important group of these effects is that during the calculation of the $\chi^2$ we took only 
the uncertainties of the Spitzer IRS spectra into account and we neglected all uncertainty related to our dust model. 

\begin{figure}[!ht]
\includegraphics[scale=0.5, angle=0]{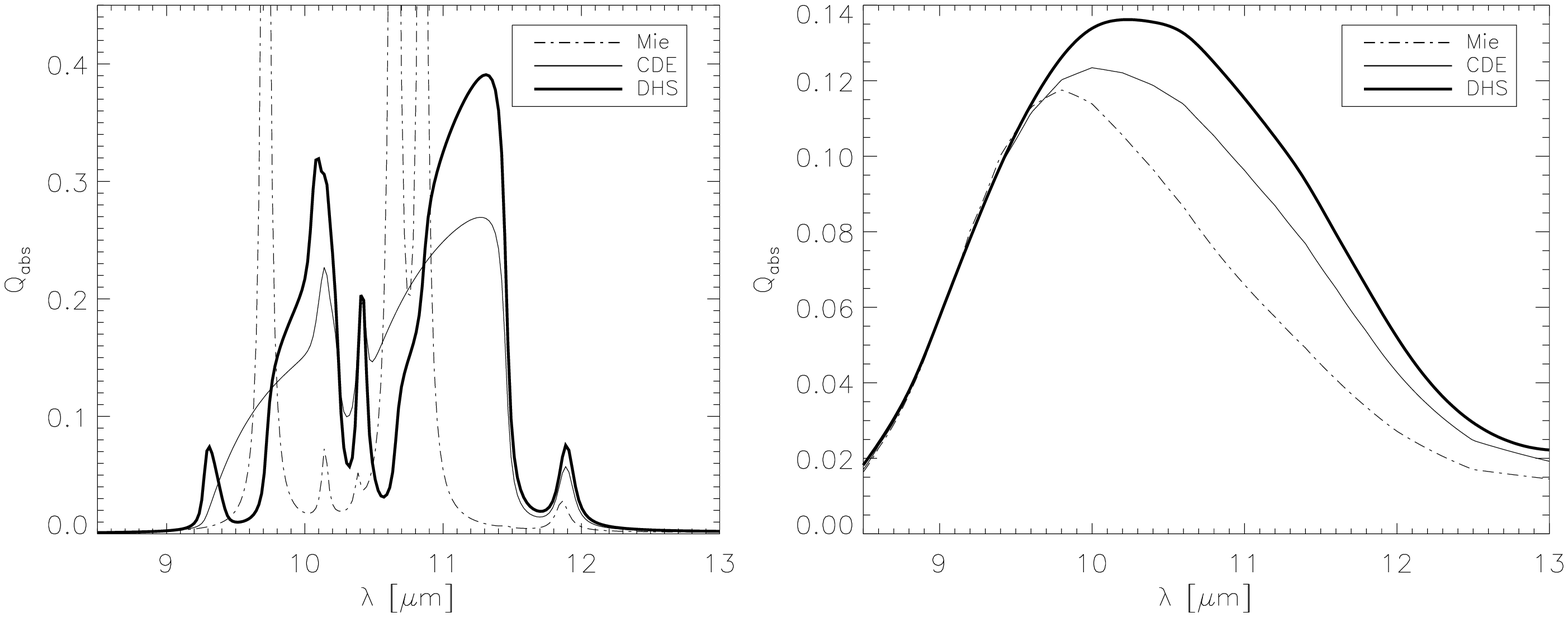}
\caption{{\it Left:} grain shape effects in the case of a 0.1\,{\micron} sized forsterite grain in the 10\,{\micron} region. 
{\it Right:} the same as for amorphous silicates with olivine stoichiometry and with Fe / (Mg+Fe) = 0.5. 
It can be seen that differences in the calculated absorption efficiencies by  DHS and CDE theories are far
smaller, than between Mie-theory and the other two scattering theories. For crystalline silicates the 
differences between the calculated optical efficiencies by arbitrary two scattering theories are larger than a 
few percent, which is a typical error
level in our fits.}
\label{fig:grain_shape}
\end{figure}

\begin{figure}[!ht]
\includegraphics[scale=0.6]{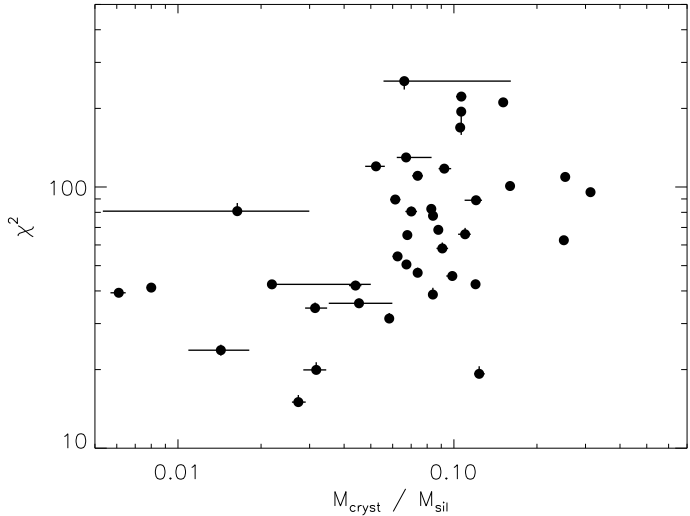}
\caption{Reduced $\chi^2$ of the fits as a function of crystallinity for the 7--17\,{\micron} region. 
In general spectra with higher crystallinity have higher reduced $\chi^2$ in the fits suggesting 
possible weaknesses in our crystalline dust model. }
\label{fig:chi_behav}
\end{figure}

For instance, it is known that grain shape is an important parameter if dust grains are in the Rayleigh domain, especially for 
crystalline silicates.  Protoplanetary dust grains are thought to have irregular shape where the calculation of the MACs from the 
optical constants are not straightforward. Differences between MACs calculated by different scattering theories in the Rayleigh 
domain are much larger than a few percent  (see Fig\,\ref{fig:grain_shape}), which is a typical discrepancy level in our fits. 
Our neglected uncertainty on the grain shape is also supported by the fact that the quality of the fit usually gets worse for 
spectra with higher crystallinity (see Figure\,\ref{fig:chi_behav}). 

Another source of uncertainty is the chemical composition of our dust model. We used the laboratory measurements of certain 
materials that are analogous to, but not necessarily the same as that in the astronomical environment. Slight differences in the chemical 
composition of the material (e.g., iron-content, Ca, Al or other ion inclusions) can already change the band profile at a percentage 
level (see Figure\,\ref{fig:opac_amorph}). Even if the composition of the material is the same, their band shapes are not necessarily
identical (see for example the measurements of Mg-rich amorphous silicates with pyroxene stoichiometry by \citet{ref:dorschner1995} 
and \citet{ref:jaeger2003}). Therefore, we certainly found the abundant \emph{types} of dust species and minerals, but we cannot claim that we found 
the \emph{exact} material composition. All these types of uncertainties are real and are present in our data/analysis, however they cannot easily be 
measured and incorporated into the calculations.  

\subsection{Amorphous silicates}
\label{sec:amorphous_silicates}

As a first step, we collected the most recent measurements of optical constants for amorphous silicates and tested the iron content
of the dust grains together with the applied scattering theory (i.e. grain shape). \citet{ref:dorschner1995} published optical constants
of glassy silicates with various iron content. Their measurements cover Fe/(Mg+Fe) ratios between 0 and 0.6 for the pyroxene 
and between 0.5 and 0.6 for the olivine family. The other set of optical constants was determined by \citet{ref:jaeger2003} on amorphous
silicates produced by the sol--gel method. In these experiments only iron-free silicates were measured. Iron-rich amorphous silicates
were not tested in our analysis, since there are no laboratory measurements of optical constants of iron rich amorphous silicates 
for both olivine and pyroxene stoichiometry available. \citet{ref:dorschner1995} measured iron-rich silicates only with olivine stoichiometry. 
 We defined three mixtures of amorphous silicates optical constants to be tested.
\begin{itemize}
\item {\bf AMIX1} iron-free silicates with optical constants from \citet{ref:jaeger2003} for the olivine and from \citet{ref:dorschner1995}
for pyroxene stoichiometry. 
\item {\bf AMIX2} iron-magnesium silicates (Fe/(Mg+Fe)=0.5) with optical constants from \citet{ref:dorschner1995}
for both olivine and pyroxene stoichiometry. 
\item {\bf AMIX3} iron-free silicates with optical constants from \citet{ref:jaeger2003} 
for both olivine and pyroxene stoichiometry. 
\end{itemize}
We calculated the MACs from the optical constants using DHS theory for a grid of f$_{\rm max}$ values, from
0 (identical to Mie theory) to 1.0. 

\begin{figure}[!ht]
\includegraphics[scale=0.6]{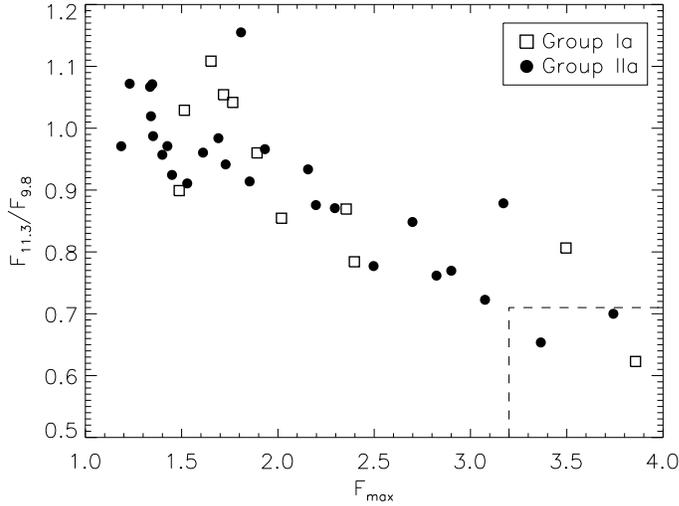}
\caption{Ratio of the normalized fluxes at 11.3\,{\micron} to that at 9.8\,{\micron}
vs. peak-to-continuum ratio of the 10\,{\micron} silicate complex. Normalized flux
was calculated as ${\rm F_\nu^{norm} = 1 + (F_\nu^{obs} - F_\nu^{cont}) / <F_\nu^{cont}>}$,
according to \citet{ref:van_boekel2005}. In the box (dashed lines)  
sources have the most pristine 10\,{\micron} silicate feature similar to that in the ISM. 
These sources were selected to test the amorphous dust population. }
\label{fig:fmax_vs_fshape}
\end{figure}

In order to study the amorphous silicates in detail we selected three sources (HD36112, HD144432 and HD152404) where the mid-infrared 
dust features show the highest possible contribution from small amorphous silicate grains over any other optically thin emission. 
In other words, (i) emission from crystalline silicates should be the lowest possible, (2) emission of amorphous grains should be 
dominated by small grains ($<1$\,{\micron}), (3) contribution of PAH emission should be the lowest possible. We use the empirical 
"feature strength vs. shape" diagram of the 10\,{\micron} silicate feature for the selection (see Figure\,\ref{fig:fmax_vs_fshape}). 
A third-order polynomial continuum is fitted to the 10\,{\micron} region for each spectrum and the feature strength is then calculated 
as F$_{\rm max}$=$1+$(F$_\nu^{\rm obs}$-F$_\nu^{\rm cont}$)/$<$F$_\nu^{\rm cont}>$ \citep{ref:van_boekel2005}, and the feature shape is the ratio of the continuum 
subtracted spectrum at 11.3\,{\micron} and 9.8\,{\micron}. Pristine 10\,{\micron} features lie in the bottom right corner of this diagram, 
while 10\,{\micron} complexes with the strongest contribution from large grains and crystalline silicates lie in the upper left corner. 
The selection criteria were F$_{\rm max}>3.2$ and F$_{11.3}$/F$_{9.8}<$0.71. 

\begin{figure}[!ht]
\includegraphics[scale=0.6, angle=0]{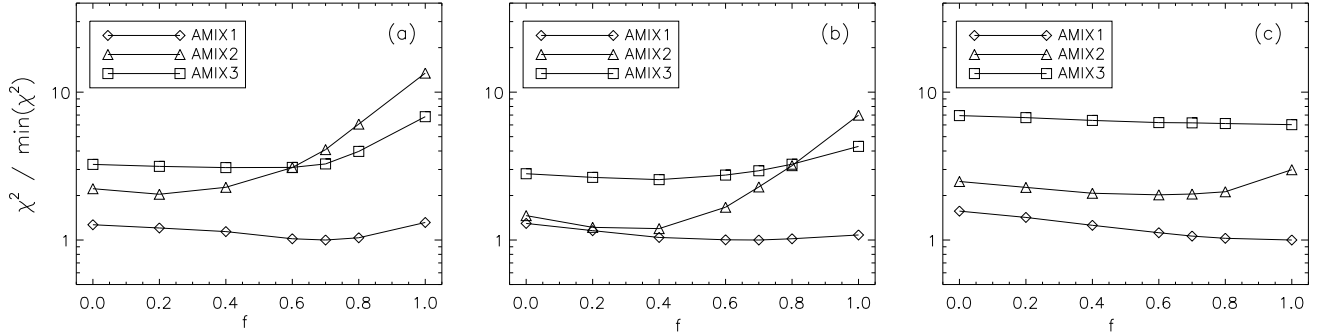}
\caption{Comparison of different datasets of amorphous silicates used for the fit. The
three panels show the results for a) HD36112, b) HD152303 and c) HD144432, respectively.
AMIX1 and AMIX3 mixtures consist of iron-free silicates while for AMIX2 Fe/(Mg+Fe)=0.5.
For the details of the different amorphous silicate mixtures, see Section\,\ref{sec:amorphous_silicates}.
The AMIX1 gives always a lower $\chi^2$ than either of the two other mixtures.}
\label{fig:comp_amsil}
\end{figure}

 Spectra of the selected sources were fitted using the MACs calculated from the AMIX1, AMIX2 and AMIX3 mixtures separately for the 
amorphous silicate components. Fits were performed for each value of the f$_{\rm max}$ grid (i.e. different grain shapes).
The MACs of the other dust components used in these fits were identical to those in the final fits and their mass fractions
were allowed to vary.
The results are summarized in Figure\,\ref{fig:comp_amsil}. It can easily be seen that the AMIX1 mixture of iron-free silicates gives always lower
$\chi^2$, than the other two mixtures. The increase in the $\chi^2$ if we used AMIX2 or AMIX3 instead of AMIX1 was always far more 
than 3\,$\sigma$. We also noted that fits with AMIX1 or AMIX3 (i.e. pure magnesium silicates) resulted in
higher value of crystallinity compared to AMIX2. For the AMIX1 mixture the best fit, i.e. lowest $\chi^2$, is obtained if one used a DHS
theory with  f$_{\rm max}=0.7$, which means that the dust grains should be porous. It is also interesting to note, that using 
iron-magnesium silicates (AMIX2) less porous grains (i.e. lower values of f$_{\rm max}$) usually give better fits than more porous
grains. During the analysis of the whole sample of stars we used the iron-free amorphous silicates of AMIX1.

 The porous iron-free amorphous silicates can reproduce the observed amorphous silicates feature better than compact iron-magnesium silicate grains. 
Although at first glance this suggests a possible lower iron-content of the amorphous silicates than it was previously thought, the conclusion is not straightforward.
If dust grains contained the iron in the form of metallic inclusions the presence of iron will not change the shape of the mid-infrared silicate bands
(e.g., \citealt{ref:ossenkopf1992}). Moreover, amorphous silicates even with the same chemical composition can show slight differences in the band shape 
(see the discussion in e.g., \citealt{ref:jaeger2003}). Although we used magnesium amorphous silicates for the spectral analysis we will not discuss the
presence or lack of iron in further details.  The fits show that the size of the amorphous grains varies between 0.4\,{\micron} and 5\,{\micron} indicating 
significant grain growth in the disk atmosphere. 

\subsection{Crystalline silicates}
\label{sec:crystalline_silicates}

Sharp bands of crystalline silicates can be observed in almost all spectra indicating that thermal processing of protoplanetary dust 
has already taken place in these disks. The value of crystallinity ranges from $\sim$1\,\% up to $\sim$30\,\%, which is about the same
range found in lower mass T Tauri stars (see e.g., \citet{ref:bouwman2008, ref:meeus2009}).  
The size of the silicate crystals, derived from the short wavelength fits, 
is found to be significantly smaller than that of the amorphous grains. In the 17--35\,{\micron} wavelength interval it is hard to determine
the size of the amorphous grains due to the very smooth mass absorption coefficient curve of the amorphous silicates. The size
of the enstatite grains is usually larger than that of the forsterite grains. 

As pointed out by \citet{ref:juhasz2009}, simple 
spectral analysis methods (like the TLTD method applied in this paper) are not capable of deriving dust parameter gradients in the
disk directly. Some spatial information on the dust composition can, however, still be derived by comparing the results of fits using 
the shorter (7--17\,{\micron}) and longer (17--35\,{\micron}) wavelength regions. By comparing the two regions we found that the 
ratio of the two crystalline dust species, enstatite and forsterite, changes. The shorter wavelength domain tends to show higher
enstatite abundance compared to forsterite, than the longer wavelength bands. 

\subsection{Correlations}
\label{sec:correlations}

We searched for correlations between global system parameters and the properties of the dust grains in order
to learn more about the mechanisms driving dust grain processing in general and to link dust and global system evolution. 
Such global parameters are the stellar parameters (luminosity, temperature), disk flaring, disk mass and the slope of the 
SED at sub-millimeter wavelengths. The disk flaring was empirically parameterized by the flux ratios at 24\,{\micron} and 
8\,{\micron}\footnote{ We note that the exact choice of the wavelengths at which the flux ratio was taken does not affect our results.}.
The disk mass and sub-millimeter slope were determined for only a part of the sample where sub-millimeter - millimeter observations were available.  
To investigate the strength of the correlation quantitatively, we calculated the Pearson's correlation coefficient. 
Pearson's correlation coefficient ($r$) measures the linear 
dependence between two variables ($x,y$) and it is calculated as 
\begin{equation}
r = \frac{n\sum x_iy_i - \sum x_i\sum_{i}y_i}
{\sqrt{n\sum x_i^2 - (\sum x_i)^2}\sqrt{n\sum y_i^2 - (\sum y_i)^2}},
\label{eq:pearsonsr}
\end{equation}
where $n$ is the number of data points. 
In order to investigate the significance of the Pearson's correlation coefficient, one can calculate the probability that the observed relation can be 
produced by a random distribution with the same sample size ($n$)
\begin{equation}
p(r,n) = \frac{2\Gamma\left(\frac{n-1}{2}\right)}{\sqrt{\pi}\Gamma\left(\frac{n-2}{2}\right)}
\int_{|r|}^1(1-u^2)^{(n-4)/2}du
\label{eq:ttest_r}
\end{equation}
(see, e.g.,, \citealt{ref:taylor1997})

We did not find any correlation between stellar parameters (e.g.,, luminosity, temperature) and the dust parameters we derived. We want to note, 
however, that there is no large spread in these parameters within our sample, thus this conclusion should be taken with care.
We also looked for possible correlation between the derived dust parameters from Spitzer IRS spectra and  disk mass or slope of 
the SED at sub-millimeter wavelengths, for those sources where sub-millimeter measurements were available.
No correlation was detected between fitted dust composition and disk mass or slope of the SED at sub-millimeter wavelengths. 
The lack of correlation suggests that the dust populations traced by mid-infrared and sub-millimeter wavelengths are not related
to each other. 

\begin{figure}[!ht]
\includegraphics[scale=0.6]{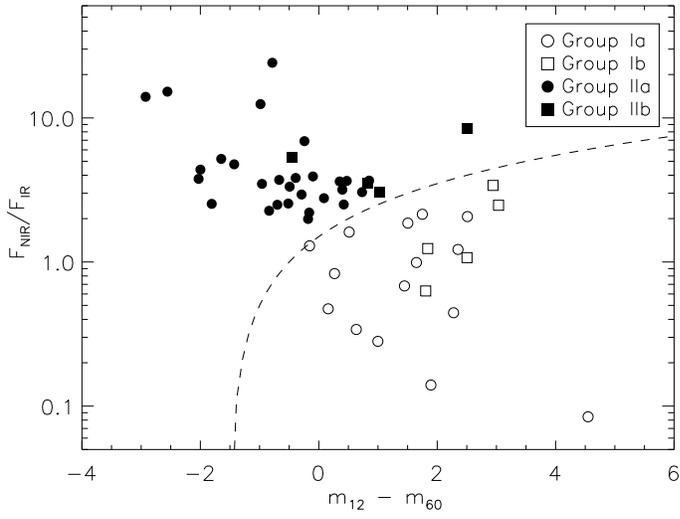}
\caption{Classification of the sources on the basis of the SED \citep{ref:van_boekel2005}. The plotted quantities are
the ratio of the near-infrared to infrared luminosities vs.
IRAS 12\,{\micron}--60\,{\micron} color (m$_{12}$ - m$_{60}$ = $-2.5\times {\rm log} F_{12}/F_{60}$). 
The dashed line marks the boundary between {\it Group I} and {\it Group II} sources, 
according to ${\rm F_{NIR} / F_{IR}}$ = $1.5\times$(m$_{12}$ - m$_{60}$).}
\label{fig:meeus_grouping}
\end{figure}

A strong correlation was found between the mass-averaged grain size of the amorphous silicates and the flaring of the disk, which
was empirically measured by the flux ratios at 24\,{\micron} and 8\,{\micron}. As can be seen in Figure\,\ref{fig:amagsize_vs_fli}, that sources 
with flatter disks have larger grains in their disk atmosphere.
We also investigated this correlation for Group\,I and Group\,II sources. For the classification of the sources we used the criterion of 
\citet{ref:van_boekel2005} based on the near-infrared to infrared luminosity ratios and the IRAS 12--60\,{\micron} color 
(see Fig.\ref{fig:meeus_grouping}). This gives the same results as the original classification scheme by \citet{ref:meeus2001}, but is 
somewhat easier to compute. Within Group\,II the Pearson's correlation coefficient is -0.65 with a probability that 
it is produced by a random distribution of $2\times10^{-4}$.  Although there are Group\,I sources which fit to the trend, defined by Group\,II sources,
there are also several outliers. Investigation of the SEDs of these outlier Group\,I sources revealed that their disk structure is clearly
different (due to e.g., the presence of a large inner hole or a gap) from that of those sources which fit to the trend. In these extreme Group\,I
sources the calculated flux ratio is likely influenced by other disk parameters (e.g., size and location of the gap or inner hole) as well, and 
the flux ratio is, therefore, not a unique and independent tracer of disk flaring. Hence, we re-calculated the Pearson's correlation coefficient for 
the sub-sample of those Group\,I and Group II sources, which have 'usual' disk structure (i.e. without a gap or larger inner hole visible in the SED).
The resulting correlation coefficient is -0.7 with a probability of $6\times10^{-6}$ that this correlation is produced by a random distribution. 

We did not detect any significant correlation between the value of crystallinity and 
any other global system parameter. However, we found that the enstatite over forsterite ratio, derived from the 5--17\,{\micron} fits differ
significantly in {\it Group I} and {\it Group II} sources, with higher values for {\it Group II} sources. The Kolmogorov-Smirnov test gives 2\% probability 
that the enstatite-forsterite ratios of the two groups originate from the same distribution.
The size of the crystals does not correlate with that of the amorphous grains, as one would expect if the crystals were produced by
annealing. In contrast to the amorphous dust population, the grain size of the crystalline silicates does not correlate with the disk flaring.
No evidence was found for correlation between amorphous grain size and crystallinity. 

\section{Discussion}
\label{sec:discussion}

\subsection{Grain growth}
\label{sec:grain_growth}

Initially the bulk of the dust in protoplanetary disks is present in sub-micron sized grains. Later on dust particles are
thought to form larger aggregates and settle to the disk midplane (e.g., \citealt{ref:dullemond_dominik2004b, ref:schraepler2004}).
By analyzing the Spitzer IRS spectra of T Tauri stars, 
\citet{ref:bouwman2008}, \citet{ref:meeus2009} and \citet{ref:sargent2009b} found a correlation between the average grain size of amorphous
silicates and the disk flaring, parameterized by the flux ratio between 30\,{\micron} and 13\,{\micron}.
The question arises whether grain growth was in fact required to produce this correlation or if dust settling without
coagulation can give the same result. \citet{ref:dullemond_dominik2008} studied this question, using two-dimensional axisymmetric
disk models. Their results suggest that the effect of dust sedimentation \emph{alone} (without coagulation)
results in stronger, sharper 10\,{\micron} silicate features in flatter disks compared to the flared ones. 
As larger dust grains settle toward the disk midplane the disk becomes geometrically flatter, 
which can be observed as a decrease of the mid- to far-infrared flux in the SED compared to near-infrared fluxes.
Since only the smallest grain population was left behind in the disk atmosphere, the emerging 10\,{\micron}
silicate feature should be strong and pointy similar to that observed in the ISM. We find the opposite trend, as the 
flux ratio between 24\,{\micron} and 8\,{\micron} decreases (flatter disks) the average grain size of the amorphous 
silicates increases. This is in agreement with the findings of \citet{ref:bouwman2008} and \citet{ref:meeus2009}
who also found the same trend in a sample of T Tauri stars. The observed correlation is, therefore, a proof that 
grain growth should have happened in these disks and that coagulation of dust particles \emph{and} dust settling 
caused the flattening of these disks. 

 The observed correlation between disk flaring and grain growth can also be enhanced
by a grain size gradient in the disk. Due to the decrease of the flaring index, the radial 
temperature (and therefore brightness) distribution in protoplanetary disks becomes steeper. This means that in flatter
disks the mid-infrared emitting region is more concentrated toward the central star than in flared disks (e.g., \citealt{ref:kessler-silacci2007}). 
If the size of the grains increases toward the central star, flatter disks show even more evolved silicate features than they would in the lack
of the grain size gradient.

\begin{figure}[!ht]
\includegraphics[angle=0,scale=0.6]{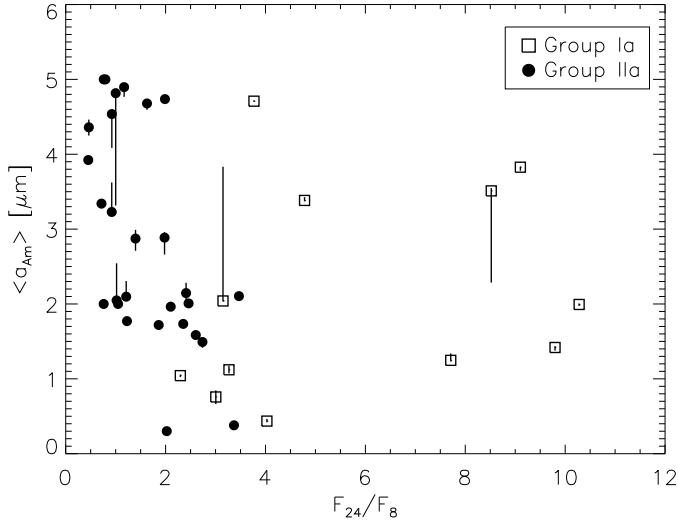}
\caption{Mass-averaged grain size vs. disk flaring. The disk flaring is empirically parameterized 
by the ratio of the flux densities at 24\,{\micron} and 8\,{\micron}. A trend is clearly visible within
{\it Group IIa}, such that sources with steeper mid-infrared SED slope have larger grains in the disk
atmosphere.  In the case of the outliers in Group\,Ia the calculated flux ratio is not likely to 
measure the disk flaring \emph{only}, but it is also influenced by other disk parameters (see the text for 
the details). }
\label{fig:amagsize_vs_fli}
\end{figure}

The found correlation also supports the evolutionary trend between Group\,I and Group\,II sources suggested by \citet{ref:dullemond_dominik2004}.
While Group\,I sources have flared disk which is dominated by small grains, the average grain size in the disk of Group\,II sources
is larger and dust settling causes the flattening of the disk. According to this prediction Group\,I sources should
occupy the bottom right regions in Figure\,\ref{fig:amagsize_vs_fli}, while Group\,II sources should be located in the top
left corner, which is indeed visible. We note again, that the outlier Group\,I sources in 
Figure\,\ref{fig:amagsize_vs_fli}, which do not fit to the main trend, have different disk structure than the rest of the sample. 
Therefore, it is not straightforward to compare them to the main trend drawn by Group\,II sources.

Although a tight correlation has been found between the disk flaring and the size of the amorphous grain population, 
no correlation was found between size of the crystals and grain size of amorphous silicates or disk flaring. 
A possible explanation for this can be if small silicate crystals are parts of larger aggregates. If the aggregate is very 
fluffy and the mass fraction of the crystalline constituents is low, crystalline bands will appear in the mass absorption
coefficients as if crystals were isolated from the large amorphous aggregate \citep{ref:min2008}.

\subsection{Silicate crystals}

It is an interesting and highly debated question how silicate crystals form in protoplanetary disks. 
So far three scenarios have been proposed for the location of crystal formation in protoplanetary disks 
and the source of heating. \citet{ref:gail2004}  proposed that silicate crystals formed 
via thermal annealing and condensation in the very inner regions of protoplanetary disks due to the heating of viscous accretion.
\citet{ref:harker_desch2002} proposed that crystalline silicates can form via thermal annealing at several 
AU from the central star if dust grains are heated by shocks. In a recent paper, \citet{ref:abraham2009} reported 
that episodic crystal formation via annealing in the surface layers of protoplanetary disks during accretion outbursts can 
also be a possible way of producing crystalline silicates.

\begin{figure}[!ht]
\includegraphics[scale=0.6]{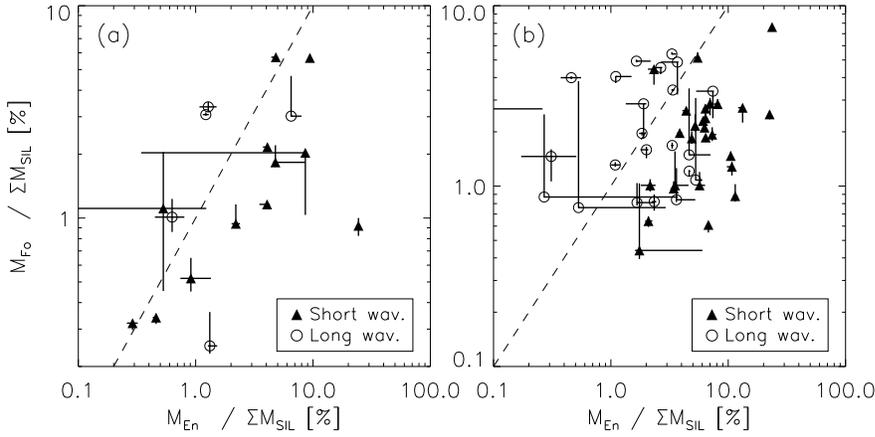}
\caption{Abundance ratios of enstatite and forsterite derived from the short
and long wavelength fits for (a) Group I and (b) for Group II sources. 
The dashed line marks the 1:1 ratio between the two plotted quantities. It can be seen
that forsterite is more abundant than enstatite at longer wavelengths (i.e. outer disk)
while at shorter wavelengths (i.e. inner disk) the situation is the opposite.  }
\label{fig:enst_vs_fors}
\end{figure}

From this study the following constraints, which should be explained by the model of crystal formation, can be identified;
\begin{itemize}
\item[] (1) Bands of crystalline silicates are seen in the 20--30\,{\micron} wavelength interval, suggesting that crystals are
also located in low temperature ($\sim$100K) regions.
\item[] (2) The forsterite-to-enstatite mass ratio changes with radius; its value is lower in the shorter wavelength fits (inner disk)
compared to that in the longer wavelengths (outer disk, see Figure\,\ref{fig:enst_vs_fors}) .
\item[] (3) Crystallinity does not correlate with any global parameter of the system (e.g., stellar parameters, disk mass, disk flaring).
\item[] (4) The size of the crystals is smaller than that of the amorphous grains.
\item[] (5) The grain size of crystalline silicates does not correlate with that of the amorphous grains.
\item[] (6) No strong evidence has been found for the correlation of crystallinity and amorphous grain size.
\item[] (7) The size of the enstatite grains is usually larger than that of the forsterite crystals.
\end{itemize}

All three models of crystal formation are consistent with (1). In the model of \citet{ref:gail2004} and \citet{ref:abraham2009}
crystals are formed in the inner disk ($<$1--2\,AU) and then transported to the outer, colder regions by large-scale mixing. In the case
of shock heating crystals are produced in the outer regions. In the models of \citet{ref:gail2004} crystal formation occurs
under equilibrium conditions. Under chemical equilibrium the crystal population is dominated by enstatite, assuming solar element 
abundances. Forsterite can only be the dominant crystalline component in the very inner disk due to its somewhat higher stability limit than 
enstatite. \citet{ref:gail2004}, therefore, predicts decreasing abundance of forsterite relative to enstatite as a function of radius in
protoplanetary disks. Our finding (2) contradicts this prediction, suggesting that the crystals we see in the spectra of HAe stars were not 
formed under equilibrium conditions. If crystals were produced by shock heating the time is probably too short to achieve chemical equilibrium. 
This has already been proved by laboratory annealing experiments \citep{ref:fabian2000}. These experiments demonstrated that annealing
of small (0.1--1\,{\micron} sized) porous silicate grains with olivine and pyroxene stoichiometry results in the formation of forsterite
regardless of the starting stoichiometry. These experiments also showed that the formation of forsterite, in the case of these small particles, 
is about an order of magnitude faster if the starting stoichiometry of the amorphous silicates is olivine instead of pyroxene. 
In the case of shock heating, where the annealing time is very short (less than an hour), one would expect the transformation of amorphous 
silicates with olivine stoichiometry to forsterite. Thus, the resulting crystal population will be dominated by forsterite while in the 
amorphous phase the abundance of grains with pyroxene stoichiometry should increase. 
The model of \citet{ref:abraham2009} does not have a clear prediction for the forsterite-to-enstatite ratio. Depending on the duration 
of the outburst and on the time a dust grain spent in the disk atmosphere the mass of the produced enstatite crystals can vary.
\citet{ref:bouwman2008} and \citet{ref:meeus2009} also speculated on the formation of enstatite. Our findings support their conclusion
that the formation of enstatite probably occurs at the very inner regions of protoplanetary disks. The temperature can be high enough
to reach chemical equilibrium during the crystal formation, favoring the formation of enstatite instead of forsterite. 
Moreover, the higher densities of the inner disk compared to the outer regions can result in frequent
collisions between dust particles making the dust grains more compact which also favors the enstatite formation \citep{ref:bouwman2008}

In the model of \citet{ref:gail2004} the required heating power for the crystallization is provided by the viscous heating of accretion in 
the disk midplane. This model, thus,  predicts a correlation between the mass of the crystals and the accretion rate. It is, however, not 
obvious if this scenario predicts any correlation between the observed crystallinity and any of the global parameters of the system. 
In this scenario, crystals form in the disk midplane, thus the amount of crystals, that can be observed in the disk atmosphere, depends
on the efficiency of both the radial and the vertical mixing in the disk, which are basically unknown factors.
Since shocks in protoplanetary disks can be produced by several different mechanisms (e.g., bow shock of a planetary embryo, see 
\citet{ref:desch2005}) and they can act at the same time, this model does not predict a strong correlation between crystallinity and global 
parameters of the system. The episodic crystal formation scenario, in agreement with (3), predicts even less correlation between crystallinity 
and any parameter of the star or the disk. Since outbursts, which provide the heating power for the crystallization, happen randomly and the total number 
of outburst a system goes through is also unpredictable, this scenario predicts a large scatter in the value of crystallinity even within 
stars with similar parameters. 

Our findings (4) and (5) can be consistent with both shock heating and midplane crystallization. If crystals were formed via condensation 
\citep{ref:gail2004} one would not expect any correlation between the sizes of the crystals and the sizes of the initially amorphous material.
 This is due to the fact that the evaporation and re-condensation of dust grains erases all correlation between the size crystals and the parent amorphous grains.
If crystal formation occurs via annealing one would expect, however, a correlation between the sizes of the starting and the resulting materials.
This correlation can only be erased if large amorphous aggregates are disrupted before or during the annealing process.
 During interactions with a shock, dust grains are not only annealed but larger aggregates are also likely to be disrupted. 
The details of the aggregate disruption depend on the physical conditions in the shock, which can vary significantly in protoplanetary disks.
Thus, shock heating predicts weak or no correlation between the size of the crystals and the amorphous grains.
In episodic crystal formation, however, it is not obvious how amorphous grains can be annealed and at the same time the grain size can also be changed. 
While shock heating and midplane crystallization naturally explain
the lack of correlation between crystallinity and amorphous grain size (6), in case of episodic surface crystallization 
still remains to be explained.

\subsection{Quality of the dust model}
\label{sec:dust_model_quality}
One of the main questions, we tried to answer in this study, is whether the dust model in protoplanetary disks
we derived from lower S/N data is sufficient to fit the spectra with high S/N ratios.
In other words, will spectra with higher S/N reveal new dust components or does the current dust model,
previously used to analyze spectra of protoplanetary disks, already contain all the main dust species which show features 
in the mid-infrared domain? In our analysis we did \emph{not} find any concluding evidence for new dust species in our
high S/N spectra. However, we found small differences between the peak positions seen in the Spitzer IRS spectra
and in our calculated absorption coefficients.  

\begin{figure}[!ht]
\includegraphics[scale=0.6]{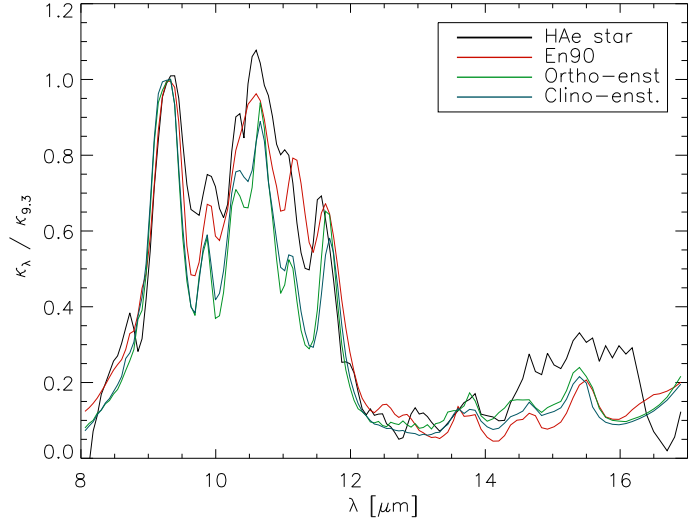}
\caption{Comparison of the MACs of enstatite derived from the Spitzer IRS spectra and
laboratory measurements \citep{ref:chihara2002}. See Section\,\ref{sec:dust_model_quality} for the details of the derivation of the 
enstatite absorption coefficients from the IRS spectra. It can be seen that the absorption coefficients derived
from the IRS spectra are the most similar to that of "En90", which is a crystalline pyroxene with about 10\,\%
iron content.}
\label{fig:enst_res}
\end{figure}

The spectrum of everal sources (e.g.,, HD190073, HD35929, HD244604) show a smooth 10\,{\micron} silicate feature, which 
does not show any substructure apart from a strong 9.3--9.4\,{\micron} peak. It is also interesting to note that in the 
spectra of these sources one can find several small peaks in the 13--16\,{\micron} wavelength interval, 
suggesting the presence of pyroxene-type crystals in the spectrum. Ortho- or clino-enstatite 
cannot reproduce the strong 9.3\,{\micron} peak and the smooth substructure-less 10\,{\micron} feature 
at the same time. We derived the crystalline pyroxene MACs from the observed 
spectra. This has been done by subtracting all other fitted components (continuum, amorphous silicate, 
forsterite, silica) from the observed spectrum and dividing the resulting curve by the underlying source function. 
We derived the MAC curve from the spectra of four sources 
(HD190073, HD244604, HD35929, HD179218), where the 9.3\,{\micron} feature was the strongest. 
The average curve is presented in Figure\,\ref{fig:enst_res}. We compared the derived MAC curve to that of 
ortho- and clino-enstatite and an iron bearing crystalline pyroxene \citep{ref:chihara2002}.
Figure\,\ref{fig:enst_res} shows that the derived MAC curve is similar to that of 
pyroxene crystals containing 10\,\% iron (denoted by En90 in \citealt{ref:chihara2002}). 
A similar comparison was done by \citet{ref:bowey2007}, who compared the peak positions of pyroxene crystals with 
different iron content to the mid-infrared spectra of HD104237 and HD179218 from \citet{ref:van_boekel2005}. On the basis of this comparison these authors suggested 
that the pyroxene crystals around these stars likely contain 10\,\%--25\,\% iron. By analyzing the Spitzer IRS spectra
of T\,Tauri stars \citet{ref:sargent2009} also arrived to the conclusion that crystalline pyroxenes are unlikely
to be pure magnesiasilicates. This implies that crystalline silicates in protoplanetary disks are not necessary iron-free 
as it is usually assumed. 

\begin{figure}[!ht]
\includegraphics[scale=0.6]{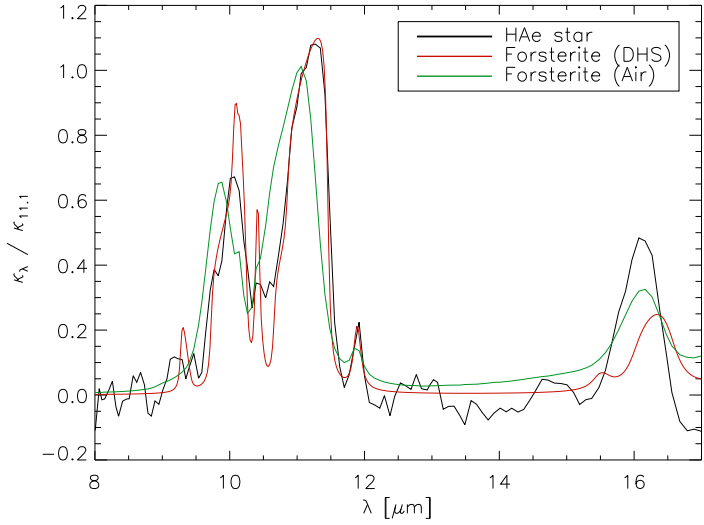}
\caption{Comparison of the MACs of forsterite derived from the Spitzer IRS spectra (see Section\,\ref{sec:dust_model_quality}), 
calculated using DHS theory (used for spectral decomposition) and measured in laboratory. In the laboratory 
experiment the MACs were measured on free flying particles \citep{ref:tamanai2009}. Although the positions of 
the bands at 10\,{\micron} and 11.3\,{\micron} are better reproduced by DHS calculations than laboratory 
measurements, in the case of the 16\,{\micron} band the situation is the opposite. Neither of the two MAC
 curve (DHS, laboratory measurement) can reproduce all the observed peak positions of
forsterite in the same time. }
\label{fig:fors_res}
\end{figure}

Almost all sources show a 16\,{\micron} feature usually associated with crystalline forsterite. Although our
dust model can fit the other forsterite bands in the 7--17\,{\micron} interval quite well, the 16\,{\micron} band was never
fitted with the same quality as the rest of the spectrum in this interval. In order to investigate this phenomenon
in detail, we derived the forsterite absorption coefficients from the spectrum of HD100546, which shows the
strongest contribution from forsterite in the 7--17\,{\micron} wavelength range. The derivation process was
similar to that applied to derive the enstatite absorption coefficients. We subtracted all other contributions
(continuum, PAHs, silica, amorphous silicates, enstatite) from the observed spectrum. The resulting 
curve is presented in Figure\,\ref{fig:fors_res}. We also compared the derived MACs
to the calculated ones which have been used for the fitting and those measured in the laboratory. For the
comparison we used the measurements of \citet{ref:tamanai2009} where the absorption coefficients were 
measured on free-flying particles instead of embedded in a KBr pellet. It can be seen that the position of the 
16\,{\micron} band is more consistent with the MACs measured on free-flying particles. On the other hand, the peak 
position of the 10.0\,{\micron} and the 11.3\,{\micron} bands are reproduced better in the  
MACs calculated using DHS theory. 
 In general the calculated MACs matched the observed bands better in the 17--35\,{\micron} wavelength interval
than in the 5--17\,{\micron} region. However, in two cases (HD244604 and HD203024) the forsterite bands also longward
of 22\,{\micron} seem to be shifted toward shorter wavelengths, similarly to the 16\,{\micron} feature.
The reason for the difference between the peak position of the calculated and the
observed forsterite bands is probably twofold. Intrinsic differences in the exact shape
of the forsterite crystals result in shifts of the band position of about 0.1\,{\micron} or even more, 
depending on the band and the shape of the particles \citep{ref:tamanai2009,ref:mutschke_2009}.
Another possible reason can be found in the scattering theory we used to calculate the MACs from the
optical constants. As pointed out by \citet{ref:mutschke_2009} calculating the absorption
cross section along each crystallographic axis independently and then taking the average can result
in the wrong band positions for anisotropic irregular particles.

\section{Conclusions}

In this paper we analyzed the Spitzer IRS spectra of a comprehensive sample of Herbig Ae stars. 
The spectra have very high S/N ratios (usually of the order of several hundred), which
allowed us not only to study dust evolution in the disks around HAe stars, but also to investigate
if the high S/N spectra reveal any new dust component. From this study we concluded, that:
\begin{itemize}
\item On the basis of the analyzed spectra, the major constituents of protoplanetary dust around Herbig Ae/Be stars are
        identified as amorphous silicates with olivine and pyroxene stoichiometry, crystalline forsterite and enstatite and silica. 
        No strong evidence for new 
	dust species has been found. However, we found slight deviations in the peak positions of the crystalline band seen
	in the Spitzer IRS spectra and in our calculated MACs. The position of the 16\,{\micron} band of 
	crystalline forsterite always peaks at shorter wavelengths than in our calculated MACs. In some cases
	the 24\,{\micron} band of forsterite is also shifted toward shorter wavelengths in the spectra compared
	to the MACs. The 8--14\,{\micron} region of several spectra shows only one narrow peak at $\sim$9.4\,{\micron}
	on top of the broad, smooth amorphous feature. We found that the 9.4\,{\micron} peak can most likely be
	associated with crystalline pyroxene with 10\% iron content. 
	
\item A tight correlation was found between the average grain size of the amorphous silicates and the flux ratios
	between 24\,{\micron} and 8\,{\micron}, used as a proxy of disk flaring. This finding is consistent with what 
	has been found by \citet{ref:bouwman2008} and \citet{ref:meeus2009} for T Tauri stars. The importance of this correlation
	is twofold. First, it is a strong observational proof that grain growth is required to produce the observed 
	diversity in grain sizes derived from mid-infrared spectroscopy. Second, it also a strict proof that 
	coagulation of dust grains and the accompanying sedimentation are responsible for the flattening of the disks
	observed as the decrease of the mid- to far-infrared radiation in disks. 

\item Those Group\,I sources which do not fit to the main trend between grain size and disk flaring were found 
        to have different disk structure than the rest of the sample. The outlier Group\,I sources have large inner holes, 
        or gaps which are visible in the SED. In these sources the calculated flux ratio between 24\,{\micron} and 8\,{\micron}
        does not only measure the flaring index, but also it is strongly affected by other disk parameters (e.g., size and location
        of the inner hole or gap). 

\item We compared the predictions of currently existing theories for crystallization to the derived dust parameters
	and the potential correlations of these parameters with each other and global system parameters. 
	We found that none of the three investigated crystallization mechanisms can \emph{alone} reproduce \emph{all}
	the observed correlations at the same time. It seems, therefore, very likely that all three processes
	may play at least some role in the evolution of protoplanetary dust particles. 

\item Crystallization in the disk midplane by accretion heating \citep{ref:gail2004} predicts increasing enstatite-to-forsterite
	ratios with radius, in contrast to what is observed in the spectra of HAe stars. This scenario also has problems reproducing
	the observed lack of correlation of crystallinity with any stellar or disk parameters. If the accretion heating was responsible
	for the heating of the amorphous grains to crystallize them, the produced mass of crystals should correlate with the accretion
	rate, which is correlated with the stellar mass. Therefore, we should have observed increasing crystallinity with increasing 
	stellar mass, which is not the case.
	
\item In crystal formation by shock heating in the outer disk (at a few tens of AUs) amorphous dust grains are heated above the annealing
	temperature only for a very short time (couple of minutes to hours) during which the crystallization should occur \citep{ref:harker_desch2002}.
	Over such a short time-scale chemical equilibrium cannot be achieved and the resulting crystal product will be forsterite
	independently of the starting stoichiometry of the amorphous particles. This can explain the observed dominance of forsterite 
	in the  outer disk, but it cannot explain the existence of enstatite at the same location. The formation of enstatite via annealing from forsterite
	is a slow process and it requires too long time to be locally produced by shock heating. 
	
\item The recently suggested episodic crystal formation in the surface layers of protoplanetary disks by \citet{ref:abraham2009} can naturally
	explain the observed diversity in the dust parameters and the lack of correlation of crystallinity with basically all the global parameters
	of the system.  However, the observed lack of correlation between crystallinity and the size of the amorphous grains and the smaller 
        size of crystals compared to amorphous grains still remains to be explained in this framework. 
\end{itemize}

\acknowledgments
We thank the anonymous referee for the careful review of the manuscript
that helped to improve the paper.

\bibliographystyle{aa}
\bibliography{ms}


\begin{deluxetable}{lllcccl}
\tablecaption{Source list of the sample and their coordinates.}
\tabletypesize{\small}
\tablewidth{14.5cm}
\tablehead{
\colhead{Name} & \colhead{Ra (J2000)}             & \colhead{Dec (J2000)}            & \colhead{Associated} &    \colhead{Spec. type} &\colhead{Group} &\colhead{Dist}\\
\colhead{}     & \colhead{[$^\circ$  $^{'}$  $^{"}$]} & \colhead{[$^\circ$  $^{'}$  $^{"}$]}& \colhead{Region} & \colhead{      }   &\colhead{       }&\colhead{[pc]}}
\startdata
 AB Aur       &    4 55 45.84$^a$   &   +30 33 04.5$^a$   & L1519       & A0Ve+sh$^c$    &     Ia    &   144.0$^a$   \\ 
 HD31648      &    4 58 46.26$^a$   &   +29 50 37.2$^a$   &             & A3ep+sh$^c$    &    IIa    &   131.0$^a$   \\ 
 HD34282      &    5 16 00.48$^a$   &    -9 48 35.4$^a$   &             & A0e$^c$        &     Ib    &   164.0$^a$   \\ 
 HD35187      &    5 24 01.17$^a$   &   +24 57 37.8$^a$   & L1559       & A2/3IV/Ve$^c$  &    IIa    &   150.0$^a$   \\ 
 HD35929      &    5 27 42.79$^a$   &    -8 19 38.4$^a$   & Orion OB1c  & F0IIIe$^c$     &    IIa    &   510.0$^j$   \\ 
 HD36112      &    5 30 27.53$^a$   &   +25 19 57.3$^a$   &             & A5IVe$^c$      &     Ia    &   204.0$^a$   \\ 
 HD244604     &    5 31 57.24$^b$   &   11 17 41.46$^b$   & Orion OB1a  & A0$^d$         &    IIa    &   336.0$^j$   \\ 
 HD36917      &    5 34 46.99$^b$   &   -5 34 14.59$^b$   & Orion OB1c  & A0V$^d$        &    IIb    &   510.0$^?$   \\
\enddata
\label{tab:source_list}
\tablerefs{(a) \citet{ref:hip1997}, (b) \citet{ref:cutri2003}, (c) \citet{ref:van_den_ancker1998}, (d)
\citet{ref:malfait1998}, (e) \citet{ref:mora2001}, (f) \citet{ref:vieira2003}, (g) \citet{ref:the1994},
(h) \citet{ref:mannings1998}, (i) \citet{ref:kun2000}, (j) \citet{ref:acke2005}, (k) \citet{ref:manoj2006},
(l) \citet{ref:wade2007}, (m) \citet{ref:hernandez2004}, (n) \citet{ref:carkner1998} (o) \citet{ref:acke2004}, 
(p) \citet{ref:hillenbrand1997}}
\end{deluxetable}

\clearpage
\begin{deluxetable}{llllll}
\tablecaption{Overview of dust species used. For each component we specify its  amorphous vs. crystalline structure, 
chemical composition, shape and reference to the 
laboratory measurements of the optical constants.}
\tabletypesize{\small}\
\tablewidth{0pt}
\tablecolumns{6}
\tablehead{\colhead{\#}&\colhead{Species}&\colhead{State}&\colhead{Chemical}&\colhead{Shape}&\colhead{Ref} \\
 \colhead{}& \colhead{}&\colhead{} & \colhead{ formula}  &  \colhead{}  & \colhead{} } \startdata
1 & Amorphous silicate  & A  &  Mg${_2}$SiO$_{4}$       & Hollow Sphere   & (1)  \\
  & (Olivine stoichiometry)&  &                     &                &  \\
2 & Amorphous silicate  & A  &  MgFeSiO$_{4}$       & Hollow Sphere   & (2)  \\
  & (Olivine stoichiometry)&  &                     &                &  \\
3 & Amorphous silicate  & A  &  MgSiO$_{3}$ & Hollow Sphere   & (1,2)  \\
 & (Pyroxene stoichiometry)&  &                     &                &  \\
4 & Amorphous silicate  & A  &  MgFeSi${_2}$O$_{6}$ & Hollow Sphere   & (2)  \\
 & (Pyroxene stoichiometry)&  &                     &                &  \\
5 & Forsterite   & C  &  Mg$_{2}$SiO$_{4}$   & Hollow Sphere & (3)   \\
6 & Ortho-Enstatite    & C  &  MgSiO$_{3}$         & Hollow Sphere & (4) \\
7 & Silica       & A  &  SiO$_{2}$           & Hollow Sphere & (5)  \\
\tableline
\enddata
\label{tab:optconst}
\tablerefs{(1) \citet{ref:jaeger2003}; (2) \citet{ref:dorschner1995} (3) \citet{ref:sogawa2006}; (4) \citet{ref:jaeger1998}; (5) \citet{ref:henning1997} }
\end{deluxetable}

\clearpage

\begin{deluxetable}{ccccl}
\tablecaption{Summary of the strongest, most frequently observed features in the Spitzer spectra of HAe stars. The identifications
were done on the basis of \citet{ref:chihara2002} and \citet{ref:koike2003} for enstatite and forsterite, respectively. The abbreviations in the case
of enstatite are identical to those used in \citet{ref:chihara2002} (CEn - clino-enstatite, OEn - ortho-enstatite, En90 - crystalline pyroxene with 10\% 
iron content). $\lambda_{\rm min}$ and $\lambda_{\rm max}$ are the shortest and longest wavelengths at which the feature is observed. $<\lambda_{\rm c}>$ is
the average peak position. }
\tablehead{\colhead{$<\lambda_{\rm c}>$} & \colhead{$\lambda_{\rm min}$} & \colhead{$\lambda_{\rm max}$} & \colhead{Nr. of sources} & \colhead{Identification}}
\startdata
      6.26  &	 6.22	&   6.28       &	 31  &    PAH					 \\ 
      7.89  &	 7.81	&   8.03       &	 21  &    PAH					 \\ 
      8.22  &	 8.08	&   8.45       &	 21  &    Silica				 \\ 
      8.67  &	 8.57	&   8.77       &	 24  &    PAH					 \\ 
      9.34  &	 9.17	&   9.49       &	 26  &    Enstatite(CEn,OEn,En90)		 \\ 
      9.91  &	 9.69	&   10.1       &	 22  &    Forsterite, Enstatite(CEn,OEn,En90)	 \\ 
      11.24  &    11.14   &   11.43	  &	    48  &    Forsterite, PAH			    \\ 
      12.68  &    12.60   &   12.84	  &	    29  &    PAH, Silica				    \\
\enddata
\label{tab:feature_list}
\end{deluxetable}

\begin{deluxetable}{rrrrrrrr}
\tablecaption{Best fit dust parameters and the reduced $\chi^2$ of the fit (5--17\,{\micron}). Listed are the derived mass fractions of each dust component in \% and
the fitted PAH fluxes in Jy.  The tabulated mass fractions (PAH fluxes) and their
uncertainties are rounded to the nearest hundredths (thousandths) place. In the case of the spectra where we used TIMMI2 spectra from \citet{ref:van_boekel2005} due to the lack of 
Spitzer short-high module we excluded the 6.2\,$\mu$m PAH band from the fitting, which is indicated by a dash (-).}   
\tabletypesize{\scriptsize}
\tablehead{ \colhead{} & \colhead{ AB Aur } & \colhead{ HD31648 } & \colhead{ HD35187 } & \colhead{ HD35929 } & \colhead{ HD36112 } & \colhead{ HD244604 } & \colhead{ HD36917 }} 
\startdata
$\chi^2$         &   80.83 &  110.38 &   41.94 &   19.26 &   54.24 &   82.45 &   89.55 \\ 
Am. Ol. 0.1\,$\mu$m & 48.98$^{+ 6.70}_{- 6.70}$ &  0.00$^{+ 0.00}_{- 0.00}$ &  0.01$^{+ 0.11}_{- 0.01}$ &  0.00$^{+ 0.00}_{- 0.00}$ & 49.40$^{+ 1.65}_{- 1.79}$ &  0.00$^{+ 0.00}_{- 0.00}$ &  0.00$^{+ 0.00}_{- 0.00}$ \\
Am. Ol. 2.0\,$\mu$m & 31.05$^{+ 4.46}_{- 4.64}$ & 41.60$^{+ 3.20}_{- 2.83}$ & 71.95$^{+ 0.81}_{- 0.78}$ & 16.41$^{+ 2.74}_{- 2.43}$ & 13.90$^{+ 1.59}_{- 1.59}$ &  0.00$^{+ 0.00}_{- 0.00}$ &  0.00$^{+ 0.00}_{- 0.00}$ \\
Am. Ol. 5.0\,$\mu$m &  0.02$^{+ 0.04}_{- 0.02}$ & 22.80$^{+ 3.50}_{- 3.94}$ &  0.00$^{+ 0.16}_{- 0.00}$ & 47.08$^{+ 4.09}_{- 3.93}$ &  0.02$^{+ 0.06}_{- 0.02}$ & 64.13$^{+ 0.44}_{- 0.43}$ & 86.81$^{+ 0.47}_{- 0.51}$ \\
Am. Py. 0.1\,$\mu$m & 13.38$^{+ 2.59}_{- 2.49}$ &  0.00$^{+ 0.00}_{- 0.00}$ &  0.00$^{+ 0.00}_{- 0.00}$ &  0.00$^{+ 0.00}_{- 0.00}$ & 24.54$^{+ 0.79}_{- 0.93}$ &  0.00$^{+ 0.00}_{- 0.00}$ &  0.03$^{+ 0.15}_{- 0.03}$ \\
Am. Py. 2.0\,$\mu$m &  0.02$^{+ 1.84}_{- 0.02}$ & 16.66$^{+ 0.62}_{- 0.86}$ &  5.80$^{+ 0.70}_{- 0.92}$ &  0.03$^{+ 0.76}_{- 0.03}$ &  2.65$^{+ 0.67}_{- 0.67}$ &  8.15$^{+ 0.40}_{- 0.29}$ &  0.00$^{+ 0.00}_{- 0.00}$ \\
Am. Py. 5.0\,$\mu$m &  0.00$^{+ 0.00}_{- 0.00}$ &  0.53$^{+ 2.01}_{- 0.50}$ &  1.90$^{+ 1.20}_{- 1.35}$ & 14.43$^{+ 1.75}_{- 1.75}$ &  0.00$^{+ 0.00}_{- 0.00}$ & 17.00$^{+ 0.61}_{- 0.74}$ &  0.00$^{+ 0.00}_{- 0.00}$ \\
Fors. 0.1\,$\mu$m &  0.43$^{+ 0.30}_{- 0.28}$ &  1.94$^{+ 0.06}_{- 0.06}$ &  1.00$^{+ 0.06}_{- 0.06}$ &  0.89$^{+ 0.06}_{- 0.07}$ &  2.15$^{+ 0.03}_{- 0.03}$ &  1.85$^{+ 0.02}_{- 0.02}$ &  0.00$^{+ 0.00}_{- 0.00}$ \\
\enddata
\label{tab:res1}
\end{deluxetable}
\clearpage

\begin{deluxetable}{rrrrrrrr}
\tablecaption{Same as Table\,\ref{tab:res1}.}
\tabletypesize{\scriptsize}
\tablehead{ \colhead{} & \colhead{ HD37258 } & \colhead{ BF Ori } & \colhead{ HD37357 } & \colhead{ HD37806 } & \colhead{ HD38120 } & \colhead{ HD250550 } & \colhead{ V590 Mon }} 
\startdata
$\chi^2$         &   45.63 &   68.54 &   31.36 &  129.77 &  119.98 &   39.36 &  194.36 \\ 
Am. Ol. 0.1\,$\mu$m &  0.00$^{+ 0.00}_{- 0.00}$ & 12.12$^{+ 1.95}_{- 2.28}$ &  0.00$^{+ 0.00}_{- 0.00}$ &  0.00$^{+ 0.00}_{- 0.00}$ & 25.41$^{+ 2.86}_{- 4.47}$ &  4.51$^{+ 0.43}_{- 0.51}$ &  0.00$^{+ 0.00}_{- 0.00}$ \\
Am. Ol. 2.0\,$\mu$m & 47.02$^{+ 0.54}_{- 0.58}$ & 38.84$^{+ 1.91}_{- 1.99}$ & 61.70$^{+ 0.50}_{- 0.52}$ &  2.21$^{+19.27}_{- 2.14}$ & 59.16$^{+ 4.77}_{- 3.18}$ & 33.27$^{+ 1.44}_{- 1.23}$ & 27.72$^{+ 0.69}_{- 0.75}$ \\
Am. Ol. 5.0\,$\mu$m &  0.00$^{+ 0.12}_{- 0.00}$ &  0.00$^{+ 0.00}_{- 0.00}$ &  0.00$^{+ 0.00}_{- 0.00}$ & 81.35$^{+ 2.43}_{-19.70}$ &  0.00$^{+ 0.00}_{- 0.00}$ & 54.10$^{+ 1.00}_{- 1.13}$ & 54.95$^{+ 0.58}_{- 0.63}$ \\
Am. Py. 0.1\,$\mu$m & 11.45$^{+ 0.20}_{- 0.18}$ &  2.73$^{+ 1.06}_{- 0.91}$ & 12.50$^{+ 0.16}_{- 0.18}$ &  0.00$^{+ 0.00}_{- 0.00}$ &  7.54$^{+ 0.37}_{- 0.37}$ &  7.08$^{+ 0.16}_{- 0.17}$ &  0.00$^{+ 0.00}_{- 0.00}$ \\
Am. Py. 2.0\,$\mu$m & 18.48$^{+ 0.79}_{- 2.15}$ & 21.73$^{+ 1.01}_{- 1.05}$ & 15.04$^{+ 0.42}_{- 0.46}$ &  3.50$^{+ 0.92}_{- 0.61}$ &  0.00$^{+ 0.00}_{- 0.00}$ &  0.15$^{+ 0.20}_{- 0.11}$ &  7.25$^{+ 0.27}_{- 0.32}$ \\
Am. Py. 5.0\,$\mu$m &  0.77$^{+ 6.04}_{- 0.75}$ &  3.15$^{+ 1.43}_{- 1.27}$ &  0.00$^{+ 0.00}_{- 0.00}$ &  3.91$^{+ 1.05}_{- 1.57}$ &  0.00$^{+ 0.00}_{- 0.00}$ &  0.00$^{+ 0.00}_{- 0.00}$ &  0.00$^{+ 0.00}_{- 0.00}$ \\
Fors. 0.1\,$\mu$m &  1.69$^{+ 0.03}_{- 0.04}$ &  2.38$^{+ 0.04}_{- 0.04}$ &  1.97$^{+ 0.05}_{- 0.04}$ &  1.10$^{+ 0.44}_{- 0.07}$ &  1.14$^{+ 0.04}_{- 0.04}$ &  0.32$^{+ 0.02}_{- 0.02}$ &  0.69$^{+ 0.03}_{- 0.03}$ \\
\enddata
\label{tab:res2}
\end{deluxetable}
\clearpage

\begin{deluxetable}{rrrrrrrr}
\tablecaption{Same as Table\,\ref{tab:res1}.}

\tabletypesize{\scriptsize}
\tablehead{ \colhead{} & \colhead{ HD50138 } & \colhead{ HD58647 } & \colhead{ HD72106 } & \colhead{ HD85567 } & \colhead{ HD95881 } & \colhead{ HD98922 } & \colhead{ HD100546 }} 
\startdata
$\chi^2$         &   88.99 &   15.01 &   95.54 &   42.40 &   42.43 &   46.95 &  168.92 \\ 
Am. Ol. 0.1\,$\mu$m &  0.00$^{+ 0.04}_{- 0.00}$ &  0.00$^{+ 0.00}_{- 0.00}$ & 44.27$^{+ 0.78}_{- 0.67}$ & 10.11$^{+ 7.32}_{-10.11}$ &  0.00$^{+ 0.00}_{- 0.00}$ &  0.00$^{+ 0.00}_{- 0.00}$ &  0.09$^{+ 0.78}_{- 0.09}$ \\
Am. Ol. 2.0\,$\mu$m & 64.00$^{+ 4.22}_{- 9.40}$ & 31.69$^{+ 1.58}_{- 1.35}$ &  0.00$^{+ 0.00}_{- 0.00}$ & 15.29$^{+11.07}_{-15.29}$ & 70.06$^{+ 0.84}_{- 1.64}$ &  0.00$^{+ 0.00}_{- 0.00}$ & 82.31$^{+ 0.51}_{- 0.76}$ \\
Am. Ol. 5.0\,$\mu$m &  7.19$^{+13.75}_{- 6.18}$ & 56.35$^{+ 1.53}_{- 1.79}$ &  0.00$^{+ 0.00}_{- 0.00}$ & 52.78$^{+39.11}_{-28.32}$ &  0.69$^{+ 5.57}_{- 0.69}$ & 89.73$^{+ 0.44}_{- 0.50}$ &  0.19$^{+ 1.28}_{- 0.19}$ \\
Am. Py. 0.1\,$\mu$m &  0.00$^{+ 0.00}_{- 0.00}$ &  0.00$^{+ 0.00}_{- 0.00}$ & 10.79$^{+ 0.44}_{- 0.45}$ &  0.77$^{+ 1.06}_{- 0.77}$ &  0.03$^{+ 0.37}_{- 0.03}$ &  0.00$^{+ 0.00}_{- 0.00}$ &  0.00$^{+ 0.00}_{- 0.00}$ \\
Am. Py. 2.0\,$\mu$m &  0.09$^{+ 0.78}_{- 0.09}$ &  0.00$^{+ 0.00}_{- 0.00}$ &  6.45$^{+ 0.51}_{- 0.55}$ &  5.99$^{+ 2.09}_{- 2.89}$ &  0.15$^{+ 0.24}_{- 0.12}$ &  0.00$^{+ 0.00}_{- 0.00}$ &  0.00$^{+ 0.00}_{- 0.00}$ \\
Am. Py. 5.0\,$\mu$m &  0.27$^{+ 1.02}_{- 0.24}$ &  0.00$^{+ 0.00}_{- 0.00}$ &  0.00$^{+ 0.00}_{- 0.00}$ &  0.00$^{+ 0.00}_{- 0.00}$ &  0.00$^{+ 0.00}_{- 0.00}$ &  0.00$^{+ 0.00}_{- 0.00}$ &  0.00$^{+ 0.00}_{- 0.00}$ \\
Fors. 0.1\,$\mu$m &  1.22$^{+ 0.11}_{- 0.11}$ &  0.64$^{+ 0.04}_{- 0.04}$ &  7.67$^{+ 0.10}_{- 0.10}$ &  0.93$^{+ 0.35}_{- 0.49}$ &  1.47$^{+ 0.09}_{- 0.08}$ &  0.58$^{+ 0.04}_{- 0.04}$ &  5.63$^{+ 0.14}_{- 0.13}$ \\
\enddata
\label{tab:res3}
\end{deluxetable}
\clearpage
\begin{deluxetable}{rrrrrrrr}
\tablecaption{Same as Table\,\ref{tab:res1}.}

\tabletypesize{\scriptsize}
\tablehead{ \colhead{} & \colhead{ HD101412 } & \colhead{ HD104237 } & \colhead{ SS73 44 } & \colhead{ HD139614 } & \colhead{ HD142666 } & \colhead{ HD142527 } & \colhead{ HD144432 }} 
\startdata
$\chi^2$         &   62.55 &  117.56 &   41.20 &   34.40 &   19.94 &  210.91 &   77.59 \\ 
Am. Ol. 0.1\,$\mu$m &  0.00$^{+ 0.00}_{- 0.00}$ &  0.00$^{+ 0.00}_{- 0.00}$ &  0.69$^{+ 0.12}_{- 0.16}$ &  0.17$^{+ 0.32}_{- 0.14}$ &  7.22$^{+ 0.97}_{- 0.93}$ & 37.14$^{+ 0.70}_{- 0.62}$ &  5.92$^{+ 3.26}_{- 2.89}$ \\
Am. Ol. 2.0\,$\mu$m & 63.05$^{+ 2.33}_{- 3.35}$ &  0.00$^{+ 0.00}_{- 0.00}$ &  0.00$^{+ 0.00}_{- 0.00}$ & 38.52$^{+ 1.11}_{- 1.06}$ & 45.20$^{+ 3.44}_{- 3.73}$ & 13.57$^{+ 0.57}_{- 0.52}$ & 42.38$^{+ 2.32}_{- 2.61}$ \\
Am. Ol. 5.0\,$\mu$m &  4.80$^{+ 4.32}_{- 2.76}$ & 73.53$^{+ 0.82}_{- 0.79}$ & 91.45$^{+ 0.23}_{- 0.21}$ & 57.33$^{+ 0.86}_{- 0.86}$ &  3.42$^{+ 3.08}_{- 3.08}$ &  0.00$^{+ 0.00}_{- 0.00}$ &  0.01$^{+ 0.07}_{- 0.01}$ \\
Am. Py. 0.1\,$\mu$m &  0.00$^{+ 0.00}_{- 0.00}$ &  0.00$^{+ 0.00}_{- 0.00}$ &  2.31$^{+ 0.06}_{- 0.06}$ &  0.00$^{+ 0.00}_{- 0.00}$ &  0.00$^{+ 0.00}_{- 0.00}$ &  0.00$^{+ 0.00}_{- 0.00}$ & 19.04$^{+ 1.01}_{- 1.05}$ \\
Am. Py. 2.0\,$\mu$m &  0.00$^{+ 0.00}_{- 0.00}$ & 10.70$^{+ 1.86}_{- 1.79}$ &  4.77$^{+ 0.18}_{- 0.20}$ &  0.00$^{+ 0.00}_{- 0.00}$ &  6.36$^{+ 1.41}_{- 1.47}$ & 22.98$^{+ 0.51}_{- 0.49}$ & 14.95$^{+ 0.59}_{- 0.63}$ \\
Am. Py. 5.0\,$\mu$m &  0.00$^{+ 0.00}_{- 0.00}$ &  3.92$^{+ 2.48}_{- 2.68}$ &  0.00$^{+ 0.00}_{- 0.00}$ &  0.00$^{+ 0.00}_{- 0.00}$ & 22.15$^{+ 2.70}_{- 2.40}$ &  0.00$^{+ 0.00}_{- 0.00}$ &  0.00$^{+ 0.10}_{- 0.00}$ \\
Fors. 0.1\,$\mu$m &  2.45$^{+ 0.13}_{- 0.13}$ &  2.00$^{+ 0.16}_{- 0.17}$ &  0.33$^{+ 0.01}_{- 0.02}$ &  0.90$^{+ 0.04}_{- 0.04}$ &  1.02$^{+ 0.08}_{- 0.08}$ &  5.70$^{+ 0.07}_{- 0.09}$ &  2.36$^{+ 0.04}_{- 0.05}$ \\
\enddata
\label{tab:res4}
\end{deluxetable}
\clearpage
\begin{deluxetable}{rrrrrrrr}
\tablecaption{Same as Table\,\ref{tab:res1}.}

\tabletypesize{\scriptsize}
\tablehead{ \colhead{} & \colhead{ HD144668 } & \colhead{ HD145263 } & \colhead{ HD150193 } & \colhead{ HD152404 } & \colhead{ 51 Oph } & \colhead{ HD163296 } & \colhead{ VV Ser }} 
\startdata
$\chi^2$         &   65.92 &   58.18 &  221.96 &   38.74 &   65.46 &   80.59 &   35.88 \\ 
Am. Ol. 0.1\,$\mu$m &  0.00$^{+ 0.00}_{- 0.00}$ &  7.51$^{+ 1.57}_{- 1.63}$ &  0.01$^{+ 0.28}_{- 0.01}$ & 53.76$^{+ 3.03}_{- 2.69}$ &  0.00$^{+ 0.00}_{- 0.00}$ &  0.00$^{+ 0.00}_{- 0.00}$ &  0.00$^{+ 0.00}_{- 0.00}$ \\
Am. Ol. 2.0\,$\mu$m & 71.33$^{+ 2.73}_{- 3.47}$ & 58.64$^{+ 3.90}_{- 4.39}$ & 56.51$^{+ 1.39}_{- 1.28}$ & 11.77$^{+ 2.26}_{- 2.45}$ & 45.33$^{+ 0.00}_{- 0.00}$ & 62.36$^{+ 1.81}_{- 2.30}$ & 21.90$^{+10.18}_{- 7.37}$ \\
Am. Ol. 5.0\,$\mu$m &  2.12$^{+ 8.57}_{- 1.88}$ &  7.35$^{+ 4.26}_{- 4.43}$ &  0.11$^{+ 0.38}_{- 0.09}$ &  0.01$^{+ 0.12}_{- 0.01}$ &  0.00$^{+ 0.00}_{- 0.00}$ &  1.34$^{+ 5.33}_{- 1.25}$ & 38.89$^{+10.40}_{-14.96}$ \\
Am. Py. 0.1\,$\mu$m &  0.00$^{+ 0.00}_{- 0.00}$ &  0.00$^{+ 0.00}_{- 0.00}$ &  0.00$^{+ 0.00}_{- 0.00}$ & 21.74$^{+ 1.11}_{- 1.20}$ &  0.00$^{+ 0.00}_{- 0.00}$ &  1.57$^{+ 0.39}_{- 0.36}$ &  0.00$^{+ 0.00}_{- 0.00}$ \\
Am. Py. 2.0\,$\mu$m &  0.02$^{+ 1.01}_{- 0.02}$ &  9.04$^{+ 1.02}_{- 0.94}$ & 20.76$^{+ 0.70}_{- 0.67}$ &  0.52$^{+ 0.72}_{- 0.43}$ &  0.00$^{+ 0.00}_{- 0.00}$ & 18.44$^{+ 0.77}_{- 0.90}$ &  0.00$^{+ 0.00}_{- 0.00}$ \\
Am. Py. 5.0\,$\mu$m &  0.02$^{+ 0.22}_{- 0.02}$ &  0.00$^{+ 0.12}_{- 0.00}$ &  0.00$^{+ 0.00}_{- 0.00}$ &  0.00$^{+ 0.00}_{- 0.00}$ & 33.60$^{+ 0.00}_{- 0.00}$ &  0.00$^{+ 0.00}_{- 0.00}$ & 24.51$^{+ 1.41}_{- 1.52}$ \\
Fors. 0.1\,$\mu$m &  2.91$^{+ 0.13}_{- 0.17}$ &  2.71$^{+ 0.13}_{- 0.13}$ &  2.88$^{+ 0.05}_{- 0.05}$ &  2.11$^{+ 0.05}_{- 0.06}$ &  0.23$^{+ 0.00}_{- 0.00}$ &  2.64$^{+ 0.07}_{- 0.10}$ &  1.18$^{+ 0.21}_{- 0.15}$ \\
\enddata
\label{tab:res5}
\end{deluxetable}
\clearpage
\begin{deluxetable}{rrrrrr}
\tablecaption{Same as Table\,\ref{tab:res1}.}

\tabletypesize{\scriptsize}
\tablehead{ \colhead{} & \colhead{ T CrA } & \colhead{ HD179218 } & \colhead{ WW Vul } & \colhead{ HD190073 } & \colhead{ HD203024 }} 
\startdata
$\chi^2$         &   23.73 &  109.27 &   50.51 &  100.87 &  254.32 \\ 
Am. Ol. 0.1\,$\mu$m & 31.15$^{+ 1.61}_{- 0.99}$ & 31.87$^{+ 1.41}_{- 1.72}$ &  0.00$^{+ 0.00}_{- 0.00}$ &  0.00$^{+ 0.00}_{- 0.00}$ &  5.53$^{+49.75}_{- 5.53}$ \\
Am. Ol. 2.0\,$\mu$m & 54.06$^{+ 1.11}_{- 1.36}$ & 32.69$^{+ 1.45}_{- 1.57}$ & 50.50$^{+ 0.29}_{- 0.28}$ & 31.16$^{+ 2.27}_{- 1.71}$ & 25.01$^{+ 3.06}_{-17.32}$ \\
Am. Ol. 5.0\,$\mu$m &  0.21$^{+ 0.54}_{- 0.17}$ &  0.39$^{+ 0.64}_{- 0.24}$ &  0.00$^{+ 0.00}_{- 0.00}$ & 20.29$^{+ 2.99}_{- 3.66}$ & 46.25$^{+ 5.14}_{-46.25}$ \\
Am. Py. 0.1\,$\mu$m &  0.00$^{+ 0.00}_{- 0.00}$ &  0.00$^{+ 0.00}_{- 0.00}$ & 18.09$^{+ 0.15}_{- 0.16}$ &  0.00$^{+ 0.00}_{- 0.00}$ &  7.39$^{+12.54}_{- 1.39}$ \\
Am. Py. 2.0\,$\mu$m & 10.06$^{+ 0.84}_{- 0.77}$ &  2.95$^{+ 0.60}_{- 0.63}$ & 13.22$^{+ 0.18}_{- 0.21}$ &  0.00$^{+ 0.00}_{- 0.00}$ &  6.85$^{+ 0.76}_{- 6.85}$ \\
Am. Py. 5.0\,$\mu$m &  0.00$^{+ 0.00}_{- 0.00}$ &  0.00$^{+ 0.00}_{- 0.00}$ &  0.01$^{+ 0.08}_{- 0.01}$ & 16.16$^{+ 1.02}_{- 0.87}$ &  0.00$^{+ 0.00}_{- 0.00}$ \\
Fors. 0.1\,$\mu$m &  0.57$^{+ 0.10}_{- 0.10}$ &  0.90$^{+ 0.09}_{- 0.09}$ &  1.45$^{+ 0.03}_{- 0.03}$ &  1.06$^{+ 0.05}_{- 0.05}$ &  1.98$^{+ 1.47}_{- 0.16}$ \\
\enddata
\label{tab:res6}
\end{deluxetable}
\clearpage

\begin{deluxetable}{rrrrrrrr}
\tablecaption{Best fit dust parameters and the reduced $\chi^2$ of the fit (17--35\,{\micron}). Listed are the derived mass fractions of each dust component in \%.}

\tabletypesize{\scriptsize}
\tablehead{ \colhead{} & \colhead{ AB Aur } & \colhead{ HD31648 } & \colhead{ HD35187 } & \colhead{ HD35929 } & \colhead{ HD36112 } & \colhead{ HD244604 } & \colhead{ HD36917 }} 
\startdata

$\chi^2$         &   58.93 &   46.29 &   31.46 &   27.89 &   49.36 &   37.61 &  187.87 \\ 
Am. Ol. 0.1\,$\mu$m & 30.00$^{+ 1.31}_{- 1.31}$ &  0.00$^{+ 0.00}_{- 0.00}$ &  0.24$^{+ 0.69}_{- 0.22}$ &  0.00$^{+ 0.00}_{- 0.00}$ &  0.45$^{+ 3.62}_{- 0.45}$ &  0.00$^{+ 0.00}_{- 0.00}$ &  0.00$^{+ 0.00}_{- 0.00}$ \\
Am. Ol. 2.0\,$\mu$m &  0.00$^{+ 0.00}_{- 0.00}$ &  0.00$^{+ 0.00}_{- 0.00}$ &  0.05$^{+ 1.20}_{- 0.05}$ & 14.10$^{+ 2.72}_{- 2.84}$ &  0.00$^{+ 0.00}_{- 0.00}$ &  0.00$^{+ 0.00}_{- 0.00}$ &  0.00$^{+ 0.00}_{- 0.00}$ \\
Am. Ol. 5.0\,$\mu$m &  0.00$^{+ 0.00}_{- 0.00}$ &  0.93$^{+ 4.90}_{- 0.93}$ & 58.43$^{+ 1.34}_{- 1.64}$ &  9.11$^{+ 1.17}_{- 3.52}$ &  0.00$^{+ 0.00}_{- 0.00}$ &  0.00$^{+ 0.00}_{- 0.00}$ &  0.00$^{+ 0.00}_{- 0.00}$ \\
Am. Py. 0.1\,$\mu$m & 64.87$^{+ 1.28}_{- 1.28}$ & 84.84$^{+ 3.56}_{- 3.03}$ &  1.67$^{+ 3.50}_{- 1.43}$ & 71.99$^{+ 2.98}_{- 3.23}$ & 90.46$^{+ 0.44}_{- 1.03}$ & 84.62$^{+ 2.53}_{- 3.50}$ & 97.16$^{+ 0.30}_{- 0.30}$ \\
Am. Py. 2.0\,$\mu$m &  0.00$^{+ 0.00}_{- 0.00}$ &  1.63$^{+ 6.51}_{- 1.63}$ & 37.17$^{+ 1.51}_{- 2.09}$ &  1.11$^{+ 3.56}_{- 1.06}$ &  0.00$^{+ 0.00}_{- 0.00}$ &  0.23$^{+11.51}_{- 0.23}$ &  0.00$^{+ 0.00}_{- 0.00}$ \\
Am. Py. 5.0\,$\mu$m &  0.00$^{+ 0.00}_{- 0.00}$ &  6.06$^{+ 3.46}_{- 3.90}$ &  0.14$^{+ 0.33}_{- 0.11}$ &  0.00$^{+ 0.00}_{- 0.00}$ &  0.00$^{+ 0.02}_{- 0.00}$ &  5.85$^{+ 3.36}_{- 2.43}$ &  0.00$^{+ 0.00}_{- 0.00}$ \\
Fors. 0.1\,$\mu$m &  0.07$^{+ 0.06}_{- 0.05}$ &  2.85$^{+ 0.12}_{- 0.08}$ &  0.00$^{+ 0.00}_{- 0.00}$ &  0.33$^{+ 0.12}_{- 0.14}$ &  1.50$^{+ 0.07}_{- 0.07}$ &  4.17$^{+ 0.12}_{- 0.11}$ &  0.00$^{+ 0.00}_{- 0.00}$ \\
\enddata
\label{tab:long_res1}
\end{deluxetable}

\begin{deluxetable}{rrrrrrrr}
\tablecaption{Same as Table\,\ref{tab:long_res1}.}

\tabletypesize{\scriptsize}
\tablehead{ \colhead{} & \colhead{ HD37258 } & \colhead{ BF Ori } & \colhead{ HD37357 } & \colhead{ HD37806 } & \colhead{ HD38120 } & \colhead{ HD250550 } & \colhead{ V590 Mon }} 
\startdata
$\chi^2$         &   21.03 &   38.37 &   10.63 &   74.21 &   38.11 &   16.20 &   12.87 \\ 
Am. Ol. 0.1\,$\mu$m &  0.00$^{+ 0.00}_{- 0.00}$ &  0.00$^{+ 0.00}_{- 0.00}$ &  0.00$^{+ 0.00}_{- 0.00}$ &  5.76$^{+ 0.87}_{- 1.02}$ & 28.04$^{+ 1.07}_{- 1.11}$ &  6.93$^{+ 5.34}_{- 4.93}$ &  0.00$^{+ 0.00}_{- 0.00}$ \\
Am. Ol. 2.0\,$\mu$m &  0.00$^{+ 0.00}_{- 0.00}$ &  0.00$^{+ 0.00}_{- 0.00}$ &  0.00$^{+ 0.00}_{- 0.00}$ &  0.00$^{+ 0.00}_{- 0.00}$ &  0.00$^{+ 0.00}_{- 0.00}$ &  0.28$^{+ 4.33}_{- 0.28}$ &  0.68$^{+ 2.98}_{- 0.66}$ \\
Am. Ol. 5.0\,$\mu$m &  0.02$^{+ 1.02}_{- 0.02}$ &  0.00$^{+ 0.00}_{- 0.00}$ &  1.59$^{+ 2.18}_{- 1.40}$ &  0.00$^{+ 0.00}_{- 0.00}$ &  0.27$^{+ 0.42}_{- 0.22}$ & 19.73$^{+ 5.19}_{- 5.85}$ & 21.47$^{+ 4.14}_{- 4.30}$ \\
Am. Py. 0.1\,$\mu$m &  0.05$^{+ 2.63}_{- 0.05}$ & 93.53$^{+ 0.28}_{- 0.30}$ & 64.91$^{+ 4.91}_{- 5.32}$ & 89.42$^{+ 1.03}_{- 0.87}$ & 30.03$^{+ 0.46}_{- 0.50}$ & 28.60$^{+ 5.16}_{- 5.82}$ & 51.53$^{+ 4.76}_{- 4.22}$ \\
Am. Py. 2.0\,$\mu$m & 95.81$^{+ 0.37}_{- 0.87}$ &  0.00$^{+ 0.00}_{- 0.00}$ & 26.35$^{+ 5.48}_{- 5.71}$ &  0.00$^{+ 0.00}_{- 0.00}$ &  0.00$^{+ 0.00}_{- 0.00}$ & 30.29$^{+ 5.36}_{- 4.21}$ & 18.83$^{+ 6.13}_{- 5.43}$ \\
Am. Py. 5.0\,$\mu$m &  0.20$^{+ 1.12}_{- 0.18}$ &  0.01$^{+ 0.12}_{- 0.01}$ &  2.75$^{+ 3.16}_{- 2.58}$ &  0.15$^{+ 0.16}_{- 0.10}$ & 40.00$^{+ 0.99}_{- 0.99}$ &  5.43$^{+ 9.30}_{- 5.01}$ &  0.00$^{+ 0.02}_{- 0.00}$ \\
Fors. 0.1\,$\mu$m &  0.82$^{+ 0.07}_{- 0.08}$ &  2.08$^{+ 0.15}_{- 0.13}$ &  0.57$^{+ 0.09}_{- 0.08}$ &  3.23$^{+ 0.08}_{- 0.08}$ &  0.00$^{+ 0.00}_{- 0.00}$ &  0.00$^{+ 0.00}_{- 0.00}$ &  2.55$^{+ 0.13}_{- 0.12}$ \\
\enddata
\label{tab:long_res2}
\end{deluxetable}

\begin{deluxetable}{rrrrrrrr}
\tablecaption{Same as Table\,\ref{tab:long_res1}.}

\tabletypesize{\scriptsize}
\tablehead{ \colhead{} & \colhead{ HD50138 } & \colhead{ HD58647 } & \colhead{ HD72106 } & \colhead{ HD85567 } & \colhead{ HD95881 } & \colhead{ HD98922 } & \colhead{ HD100546 }} 
\startdata
$\chi^2$         &  106.34 &   38.52 &   42.33 &   78.85 &  152.74 &  180.41 &   67.64 \\ 
Am. Ol. 0.1\,$\mu$m & 37.14$^{+ 0.89}_{- 0.82}$ &  3.05$^{+22.19}_{- 3.03}$ &  8.87$^{+ 1.84}_{- 1.44}$ & 44.87$^{+ 0.73}_{- 0.67}$ &  0.00$^{+ 0.03}_{- 0.00}$ & 52.71$^{+ 3.39}_{- 1.74}$ &  0.00$^{+ 0.00}_{- 0.00}$ \\
Am. Ol. 2.0\,$\mu$m &  0.00$^{+ 0.00}_{- 0.00}$ & 21.39$^{+ 4.64}_{- 9.00}$ &  0.00$^{+ 0.00}_{- 0.00}$ &  0.00$^{+ 0.00}_{- 0.00}$ &  0.07$^{+ 0.73}_{- 0.07}$ &  0.00$^{+ 0.00}_{- 0.00}$ &  0.00$^{+ 0.00}_{- 0.00}$ \\
Am. Ol. 5.0\,$\mu$m &  1.66$^{+ 0.77}_{- 0.81}$ &  1.36$^{+21.24}_{- 1.36}$ &  7.38$^{+ 1.02}_{- 1.25}$ &  0.00$^{+ 0.00}_{- 0.00}$ &  0.00$^{+ 0.00}_{- 0.00}$ &  0.00$^{+ 0.00}_{- 0.00}$ &  0.00$^{+ 0.00}_{- 0.00}$ \\
Am. Py. 0.1\,$\mu$m & 52.67$^{+ 0.47}_{- 0.51}$ & 43.89$^{+10.07}_{-10.48}$ & 79.93$^{+ 0.81}_{- 0.84}$ & 48.05$^{+ 0.65}_{- 0.70}$ & 88.75$^{+ 0.15}_{- 0.28}$ & 41.06$^{+ 1.68}_{- 3.25}$ & 91.78$^{+ 0.17}_{- 0.24}$ \\
Am. Py. 2.0\,$\mu$m &  0.00$^{+ 0.00}_{- 0.00}$ & 21.17$^{+ 8.82}_{- 7.83}$ &  0.00$^{+ 0.00}_{- 0.00}$ &  0.00$^{+ 0.00}_{- 0.00}$ &  0.00$^{+ 0.00}_{- 0.00}$ &  0.00$^{+ 0.00}_{- 0.00}$ &  0.00$^{+ 0.00}_{- 0.00}$ \\
Am. Py. 5.0\,$\mu$m &  0.02$^{+ 0.18}_{- 0.02}$ &  0.00$^{+ 0.02}_{- 0.00}$ &  0.00$^{+ 0.00}_{- 0.00}$ &  0.00$^{+ 0.00}_{- 0.00}$ &  0.00$^{+ 0.00}_{- 0.00}$ &  0.00$^{+ 0.00}_{- 0.00}$ &  0.04$^{+ 0.25}_{- 0.03}$ \\
Fors. 0.1\,$\mu$m &  2.10$^{+ 0.04}_{- 0.04}$ &  0.33$^{+ 0.12}_{- 0.12}$ &  1.95$^{+ 0.04}_{- 0.05}$ &  1.17$^{+ 0.06}_{- 0.05}$ &  3.43$^{+ 0.05}_{- 0.07}$ &  1.17$^{+ 0.04}_{- 0.07}$ &  4.21$^{+ 0.15}_{- 0.11}$ \\
\enddata
\label{tab:long_res3}
\end{deluxetable}

\begin{deluxetable}{rrrrrrrr}
\tablecaption{Same as Table\,\ref{tab:long_res1}.}

\tabletypesize{\scriptsize}
\tablehead{ \colhead{} & \colhead{ HD101412 } & \colhead{ HD104237 } & \colhead{ SS73 44 } & \colhead{ HD139614 } & \colhead{ HD142666 } & \colhead{ HD142527 } & \colhead{ HD144432 }} 
\startdata
$\chi^2$         &   94.79 &   73.75 &   89.14 &   26.95 &   26.87 &  701.54 &   21.25 \\ 
Am. Ol. 0.1\,$\mu$m &  0.00$^{+ 0.00}_{- 0.00}$ &  0.00$^{+ 0.00}_{- 0.00}$ & 54.94$^{+ 0.59}_{- 0.54}$ & 17.25$^{+ 8.03}_{-14.91}$ &  0.06$^{+ 1.85}_{- 0.06}$ &  0.00$^{+ 0.00}_{- 0.00}$ &  0.05$^{+ 0.98}_{- 0.05}$ \\
Am. Ol. 2.0\,$\mu$m &  0.00$^{+ 0.00}_{- 0.00}$ &  0.00$^{+ 0.00}_{- 0.00}$ &  0.00$^{+ 0.00}_{- 0.00}$ &  8.47$^{+14.20}_{- 7.65}$ &  0.11$^{+ 1.93}_{- 0.10}$ &  0.00$^{+ 0.00}_{- 0.00}$ &  0.56$^{+ 2.80}_{- 0.53}$ \\
Am. Ol. 5.0\,$\mu$m &  0.00$^{+ 0.00}_{- 0.00}$ &  0.00$^{+ 0.00}_{- 0.00}$ &  0.00$^{+ 0.00}_{- 0.00}$ &  0.12$^{+ 0.83}_{- 0.11}$ &  0.20$^{+ 3.10}_{- 0.20}$ &  0.00$^{+ 0.00}_{- 0.00}$ &  0.08$^{+ 1.47}_{- 0.08}$ \\
Am. Py. 0.1\,$\mu$m &  0.00$^{+ 0.00}_{- 0.00}$ & 90.36$^{+ 0.16}_{- 0.15}$ & 43.57$^{+ 0.58}_{- 0.56}$ &  0.79$^{+ 3.53}_{- 0.77}$ & 32.32$^{+ 6.63}_{- 6.12}$ & 93.26$^{+ 0.23}_{- 0.29}$ & 95.49$^{+ 0.60}_{- 2.72}$ \\
Am. Py. 2.0\,$\mu$m & 88.17$^{+ 0.21}_{- 0.20}$ &  0.00$^{+ 0.00}_{- 0.00}$ &  0.00$^{+ 0.00}_{- 0.00}$ & 69.39$^{+ 2.01}_{- 1.65}$ & 59.74$^{+ 6.70}_{- 6.98}$ &  0.00$^{+ 0.00}_{- 0.00}$ &  0.00$^{+ 0.00}_{- 0.00}$ \\
Am. Py. 5.0\,$\mu$m &  0.01$^{+ 0.03}_{- 0.01}$ &  0.00$^{+ 0.00}_{- 0.00}$ &  0.08$^{+ 0.31}_{- 0.08}$ &  0.00$^{+ 0.15}_{- 0.00}$ &  0.00$^{+ 0.16}_{- 0.00}$ &  0.10$^{+ 0.14}_{- 0.09}$ &  0.04$^{+ 0.21}_{- 0.03}$ \\
Fors. 0.1\,$\mu$m &  0.83$^{+ 0.09}_{- 0.08}$ &  3.74$^{+ 0.07}_{- 0.06}$ &  0.00$^{+ 0.00}_{- 0.00}$ &  0.00$^{+ 0.00}_{- 0.00}$ &  0.63$^{+ 0.13}_{- 0.14}$ &  1.60$^{+ 0.09}_{- 0.09}$ &  0.98$^{+ 0.08}_{- 0.09}$ \\
\enddata
\label{tab:long_res4}
\end{deluxetable}

\begin{deluxetable}{rrrrrrrr}
\tablecaption{Same as Table\,\ref{tab:long_res1}.}

\tabletypesize{\scriptsize}
\tablehead{ \colhead{} & \colhead{ HD144668 } & \colhead{ HD145263 } & \colhead{ HD150193 } & \colhead{ HD152404 } & \colhead{ 51 Oph } & \colhead{ HD163296 } & \colhead{ VV Ser }} 
\startdata
$\chi^2$         &   63.15 &   30.21 &   94.90 &    6.79 &  101.24 &   31.49 &   82.00 \\ 
Am. Ol. 0.1\,$\mu$m & 25.94$^{+ 0.88}_{- 0.88}$ &  0.00$^{+ 0.00}_{- 0.00}$ &  0.00$^{+ 0.00}_{- 0.00}$ & 18.66$^{+18.12}_{-17.41}$ &  0.00$^{+ 0.00}_{- 0.00}$ &  0.00$^{+ 0.00}_{- 0.00}$ & 51.49$^{+ 0.94}_{- 0.98}$ \\
Am. Ol. 2.0\,$\mu$m &  0.00$^{+ 0.00}_{- 0.00}$ &  0.00$^{+ 0.00}_{- 0.00}$ &  0.00$^{+ 0.00}_{- 0.00}$ & 20.31$^{+20.10}_{-20.10}$ & 23.33$^{+ 1.50}_{- 1.38}$ &  0.00$^{+ 0.01}_{- 0.00}$ &  0.00$^{+ 0.00}_{- 0.00}$ \\
Am. Ol. 5.0\,$\mu$m &  0.00$^{+ 0.00}_{- 0.00}$ &  0.00$^{+ 0.00}_{- 0.00}$ &  0.00$^{+ 0.00}_{- 0.00}$ &  0.00$^{+ 0.00}_{- 0.00}$ &  0.00$^{+ 0.04}_{- 0.00}$ &  0.31$^{+ 2.26}_{- 0.31}$ &  0.27$^{+ 1.31}_{- 0.27}$ \\
Am. Py. 0.1\,$\mu$m & 70.63$^{+ 0.89}_{- 0.85}$ &  0.01$^{+ 0.04}_{- 0.01}$ & 68.68$^{+ 0.96}_{- 0.88}$ & 46.40$^{+10.58}_{-22.49}$ & 69.40$^{+ 1.37}_{- 1.37}$ & 64.94$^{+ 3.58}_{- 3.44}$ & 31.93$^{+ 1.22}_{- 1.08}$ \\
Am. Py. 2.0\,$\mu$m &  0.00$^{+ 0.00}_{- 0.00}$ & 81.29$^{+ 8.47}_{-12.71}$ & 18.73$^{+ 0.91}_{- 0.95}$ &  6.80$^{+14.27}_{- 6.72}$ &  0.00$^{+ 0.00}_{- 0.00}$ & 28.86$^{+ 3.53}_{- 3.13}$ & 11.07$^{+ 1.30}_{- 1.59}$ \\
Am. Py. 5.0\,$\mu$m &  0.14$^{+ 0.49}_{- 0.13}$ & 10.99$^{+14.29}_{- 8.04}$ &  0.01$^{+ 0.04}_{- 0.01}$ &  0.00$^{+ 0.00}_{- 0.00}$ &  0.00$^{+ 0.00}_{- 0.00}$ &  0.00$^{+ 0.00}_{- 0.00}$ &  0.00$^{+ 0.00}_{- 0.00}$ \\
Fors. 0.1\,$\mu$m &  1.34$^{+ 0.04}_{- 0.04}$ &  1.74$^{+ 0.21}_{- 0.22}$ &  3.06$^{+ 0.05}_{- 0.05}$ &  0.46$^{+ 0.15}_{- 0.16}$ &  0.00$^{+ 0.00}_{- 0.00}$ &  1.26$^{+ 0.12}_{- 0.12}$ &  0.00$^{+ 0.04}_{- 0.00}$ \\
\enddata
\label{tab:long_res5}
\end{deluxetable}

\begin{deluxetable}{rrrrrrrr}
\tablecaption{Same as Table\,\ref{tab:long_res1}.}

\tabletypesize{\scriptsize}
\tablehead{ \colhead{} & \colhead{ T CrA } & \colhead{ HD179218 } & \colhead{ WW Vul } & \colhead{ HD190073 } & \colhead{ HD203024 }} 
\startdata
$\chi^2$         &   15.84 &  651.54 &   69.70 &   83.84 &  197.33 \\ 
Am. Ol. 0.1\,$\mu$m &  0.00$^{+ 0.01}_{- 0.00}$ &  0.00$^{+ 0.00}_{- 0.00}$ &  0.00$^{+ 0.00}_{- 0.00}$ &  0.00$^{+ 0.00}_{- 0.00}$ &  0.00$^{+ 0.00}_{- 0.00}$ \\
Am. Ol. 2.0\,$\mu$m &  0.00$^{+ 0.00}_{- 0.00}$ &  0.00$^{+ 0.00}_{- 0.00}$ &  0.00$^{+ 0.00}_{- 0.00}$ &  0.00$^{+ 0.00}_{- 0.00}$ &  0.00$^{+ 0.00}_{- 0.00}$ \\
Am. Ol. 5.0\,$\mu$m &  0.00$^{+ 0.00}_{- 0.00}$ & 20.63$^{+ 0.89}_{- 1.33}$ &  0.00$^{+ 0.00}_{- 0.00}$ &  0.00$^{+ 0.00}_{- 0.00}$ &  0.00$^{+ 0.00}_{- 0.00}$ \\
Am. Py. 0.1\,$\mu$m & 35.71$^{+ 2.36}_{- 2.09}$ & 66.74$^{+ 1.26}_{- 0.91}$ &  9.05$^{+ 3.88}_{- 3.73}$ & 93.76$^{+ 0.23}_{- 0.18}$ & 86.18$^{+ 1.15}_{- 1.29}$ \\
Am. Py. 2.0\,$\mu$m & 48.69$^{+ 2.70}_{- 2.60}$ &  0.00$^{+ 0.00}_{- 0.00}$ & 81.25$^{+ 4.27}_{- 4.82}$ &  0.00$^{+ 0.00}_{- 0.00}$ &  0.17$^{+ 0.51}_{- 0.12}$ \\
Am. Py. 5.0\,$\mu$m &  6.66$^{+ 2.81}_{- 2.49}$ &  0.00$^{+ 0.00}_{- 0.00}$ &  0.00$^{+ 0.00}_{- 0.00}$ &  0.00$^{+ 0.02}_{- 0.00}$ &  8.01$^{+ 1.38}_{- 1.04}$ \\
Fors. 0.1\,$\mu$m &  0.17$^{+ 0.05}_{- 0.05}$ &  4.67$^{+ 0.03}_{- 0.04}$ &  0.82$^{+ 0.12}_{- 0.11}$ &  2.45$^{+ 0.07}_{- 0.10}$ &  3.13$^{+ 0.04}_{- 0.05}$ \\
\enddata
\label{tab:long_res6}
\end{deluxetable}

\clearpage

\vskip 2.5cm

\clearpage

\clearpage

\clearpage

\clearpage

\clearpage

\clearpage

\clearpage

\end{document}